%% file: main.tex
\documentclass[letterpaper,twocolumn,10pt,dvipsnames]{article}
\usepackage{usenix-2020-09}

\usepackage{amsmath, amssymb, amsfonts}
\usepackage{booktabs}
\usepackage{multirow}
\usepackage{subcaption}
\usepackage[nolist]{acronym}
\usepackage{xspace}
\usepackage{tikz}
\usepackage{enumitem} 
\usepackage{textcomp} 

\begin{document}

\input{sections/misc/acronyms}
\input{sections/misc/commands}
\input{sections/misc/math}

\date{}
\title{\Large \bf \tool: Taming Audio Adversarial Examples}

\author{
{\rm Thorsten Eisenhofer}\\
Ruhr University Bochum
\and
{\rm Lea Sch\"onherr}\\
Ruhr University Bochum
\and
 {\rm Joel Frank}\\
Ruhr University Bochum
\vspace{1em}
\and
 {\rm Lars Speckemeier}\\
University College London
\and
 {\rm Dorothea Kolossa}\\
Ruhr University Bochum
\and
 {\rm Thorsten Holz}\\
Ruhr University Bochum
}

\maketitle

\begin{abstract}
Adversarial examples seem to be inevitable. These specifically crafted inputs allow attackers to arbitrarily manipulate machine learning systems. Even worse, they often seem harmless to human observers. In our digital society, this poses a significant threat. For example, \emph{Automatic Speech Recognition}~(ASR) systems, which serve as hands-free interfaces to many kinds of systems, can be attacked with inputs incomprehensible for human listeners. The research community has unsuccessfully tried several approaches to tackle this problem.

In this paper we propose a different perspective: We accept the presence of adversarial examples against ASR systems, but we require them to be \emph{perceivable} by human listeners. By applying the principles of \emph{psychoacoustics}, we can remove semantically irrelevant information from the ASR input and train a model that resembles human perception more closely. We implement our idea in a tool named \tool\footnote{The French word for \emph{tamer}} and demonstrate that our augmented system, in contrast to an unmodified baseline, successfully focuses on perceptible ranges of the input signal. This change forces adversarial examples into the audible range, while using minimal computational overhead and preserving benign performance.
To evaluate our approach, we construct an \textit{adaptive attacker} that actively tries to avoid our augmentations and demonstrate that adversarial examples from this attacker remain clearly perceivable. Finally, we substantiate our claims by performing a hearing test with crowd-sourced human listeners.
\end{abstract}

\input{sections/01_introduction.tex}
\input{sections/02_background.tex}
\input{sections/03_approach.tex}
\input{sections/04_results.tex}
\input{sections/05_user_study}
\input{sections/06_related.tex}
\input{sections/07_discussion.tex}
\input{sections/08_conclusion.tex}

\paragraph{Acknowledgments}
We would like to thank our shepherd Xiaoyu Ji and the anonymous reviewers for their valuable comments and suggestions. We also thank our colleagues Nils Bars, Merlin Chlosta, Sina D\"aubener, Asja Fischer, Jan Freiwald, Moritz Schl\"ogel, Steffen Zeiler for their feedback and fruitful discussions. 
This work was supported by the Deutsche Forschungsgemeinschaft (DFG, German Research Foundation) under Germany's Excellence Strategy -- EXC-2092  \textsc{CaSa} -- 390781972.

\clearpage
\bibliographystyle{unsrt}
\bibliography{strings, references}

\appendix
\input{sections/09_appendix.tex}

\end{document}

%% file: sections/misc/acronyms.tex
\begin{acronym}
    \acro{ASR}{\emph{Automatic Speech Recognition}}
    \acro{DNN}{\emph{Deep Neural Networks}}
    \acro{IIR}{\emph{infinite impulse response}}
    \acroplural{IIR}[IIRS]{\emph{infinite impulse responses}}
    \acro{RNN}{\emph{Recurrent Neural Networks}}
    \acro{DNN-HMM}{\emph{Deep Neural  Network - Hidden  Markov  Model}}
    \acro{HMM}{\emph{hidden Markov model}}
    \acro{DFT}{\emph{discrete Fourier transform}}
    \acro{WER}{\emph{Word Error Rate}}
    \acro{STFT}{\emph{short-time Fourier transform}}
    \acro{SNRseg}{\emph{Segmental Signal-to-Noise Ratio}}
    \acro{SNR}{\emph{Signal-to-Noise Ratio}}
    \acro{WSJ}{\emph{Wall Street Journal}}
    \acro{CTC}{\emph{Connectionist Temporal Classification}}
    \acro{GMM}{\emph{Gaussian mixture model}}
    \acro{VBR}{\emph{variable bitrate}}
    \acro{CBR}{\emph{constant bitrate}}
    \acro{MUSHRA}{\emph{Multiple Stimuli with Hidden Reference and Anchor}}
\end{acronym}

%% file: sections/misc/commands.tex
\newcommand{\eg}{e.\,g.,\xspace}
\newcommand{\ie}{i.\,e.,\xspace}
\newcommand{\kaldi}{\textsc{Kaldi}\xspace}
\newcommand{\deepspeech}{\textsc{DeepSpeech}\xspace}
\newcommand{\tool}{\textsc{Dompteur}\xspace}

\newcommand{\wrt}{w.\@\,r.\@\,t.\@\xspace}
\newcommand{\etc}{etc.\@\xspace}
\newcommand{\cf}{cf.\@\xspace}
\newcommand{\etal}{et~al.\@\xspace}

\newcommand{\astretch}[1]{\renewcommand{\arraystretch}{#1}}
\newcommand{\textapprox}{\raisebox{0.5ex}{\texttildelow}}

%% file: sections/misc/math.tex
\newcommand{\timesignal}{x}
\newcommand{\filteredsignal}{y}
\newcommand{\timeindex}{t}
\newcommand{\timemax}{T}
\newcommand{\frameindex}{n}
\newcommand{\framemax}{N}
\newcommand{\freqindex}{k}
\newcommand{\frequmax}{K}

\newcommand{\complexsignal}{\mathbf{S}}
\newcommand{\maxsignal}{\underset{\frameindex,\freqindex}{\text{max}}(|\complexsignal|)}
\newcommand{\modifiedsignal}{\mathbf{N}}
\newcommand{\mask}{\mathbf{M}}
\newcommand{\diffsignal}{\mathbf{D}}
\newcommand{\marginthresholds}{\Phi}
\newcommand{\hearingthresholds}{\mathbf{H}}
\newcommand{\inputsignal}{\mathbf{T}}

\newcommand{\margin}{\Phi}

\newcommand{\complexangle}{\varphi}

%% file: sections/01_introduction.tex
\section{Introduction}
\label{sec:introduction}
The advent of deep learning has changed our digital society.
Starting from simple recommendation techniques~\cite{pazzani-07-content} or image recognition applications~\cite{krizhevsky-12-imagenet}, machine-learning systems have evolved to solve and play games on par with humans~\cite{mnih-15-human, silver-16-mastering, berner-19-dota, vinyals-19-grandmaster}, to predict protein structures~\cite{senior-20-improved}, identify faces~\cite{taigman-14-deepface}, or recognize speech at the level of human listeners~\cite{xiong-16-achieving}.
These systems are now virtually ubiquitous and are being granted access to critical and sensitive parts of our daily lives.
They serve as our personal assistants~\cite{ram-17-conversational}, unlock our smart homes' doors~\cite{misc-16-alexa}, or drive our autonomous cars~\cite{misc-19-teslaai}.

Given these circumstances, the discovery of \textit{adversarial examples}~\cite{szegedy-13-intriguing} has had a shattering impact.
These specifically crafted inputs can completely mislead machine learning-based systems.
Mainly studied for image recognition~\cite{szegedy-13-intriguing}, in this work, we study how adversarial examples can affect \ac{ASR} systems.
Preliminary research has already transferred adversarial attacks to the audio domain~\cite{song-17-inaudible, zhang-17-dolphinattack,  yuan-18-commandersong, schoenherr-19-psychoacoustics, schoenherr-20-imperio, abdullah-20-sok}.
The most advanced attacks start from a harmless input signal and change the model's prediction towards a target transcription while simultaneously \emph{hiding} their malicious intent in the inaudible audio spectrum. 

To address such attacks, the research community has developed various defense mechanisms~\cite{papernot-15-limitations, moosavidezfooli-15-deepfool, carlini-17-robustness, metzen-17-detecting, feinman-17-detecting, carlini-17-adversarial}.
All of the proposed defenses---in the ever-lasting cat-and-mouse game between attackers and defenders---have  subsequently been broken~\cite{gilmer-18-motivating}.
Recently, Shamir \etal{}~\cite{shamir-19-simple} even demonstrated that, given certain constraints, we can expect to always find adversarial examples for our models.

Considering these circumstances, we ask the following research question: \emph{When we accept that adversarial examples exist, what else can we do?}
We propose a paradigm shift: Instead of preventing \emph{all} adversarial examples, we accept the presence of \emph{some}, but we want them to be audibly changed.

To achieve this shift, we take inspiration from the machine learning community, which sheds a different light on adversarial examples:
Illyas \etal{}~\cite{ilyas-19-adversarial} interpret the presence of adversarial examples as a disconnection between human expectations and the reality of a mathematical function trained to minimize an objective.
We tend to think that machine learning models must learn meaningful features, e.\,g., a cat has paws.
However, this is a human's perspective on what makes a cat a cat.
Machine learning systems instead use \emph{any} available feature they can incorporate in their decision process.
Consequently, Illyas~\etal{} demonstrate that image classifiers utilize so-called \emph{brittle features}, which are highly predictive, yet not recognizable by humans.

Recognizing this mismatch between human expectations and the inner workings of machine learning systems, we propose a novel design principle for \ac{ASR} system inspired by the human auditory system. Our approach is based on two key insights: (i) the human voice frequency is limited to the band ranges of approximately $300-5000\,Hz$~\cite{monson-14-perceptual}, while \ac{ASR} systems are typically trained on 16kHz signals, which range from $0-8000\,Hz$, and (ii) audio signal can carry information, inaudible to humans~\cite{zhang-17-dolphinattack}. Given these insights, we modify the \ac{ASR} system by restricting its access to frequencies and applying psychoacoustic modeling to remove \emph{inaudible} ranges.
The effects are twofold:
The \ac{ASR} system can learn a better approximation of the human perception during training (i.e., discarding unnecessary information), while simultaneously, adversaries are forced to place any adversarial perturbation into audible ranges.

We implement these principles in a prototype we call \tool.
In a series of experiments, we demonstrate that our prototype more closely models the human auditory system. More specifically, we successfully show that our \acs{ASR} system, in contrast to an unmodified baseline, focuses on perceptible ranges of the audio signal.
Following Carlini~\etal{}~~\cite{carlini-19-evaluating}, we depart from the lab settings predominantly studied in prior work:
We assume a white-box attacker with real-world capabilities, i.e., we grant them full knowledge of the system and they can introduce an unbounded amount of perturbations.
Even under these conditions, we are able to force the attacker to produce adversarial examples with an average of 24.33\,dB of added perturbations while remaining accurate for benign inputs.
Additionally, we conduct a large scale user study with 355 participants.
The study confirms that the adversarial examples constructed for \tool are easily distinguishable from benign audio samples and adversarial examples constructed for the baseline system.

\smallskip\noindent
In summary, we make the following key contributions:
\begin{itemize}
    \item \textbf{Constructing an Augmented \acs{ASR}.} 
    We utilize our key insights to bring \acs{ASR} systems in better alignment with human expectations and demonstrate that traditional \acs{ASR} systems indeed utilize non-audible signals that are not recognizable by humans. 

    \item \textbf{Evaluation Against Adaptive Attacker.} 
    We construct a realistic scenario where the attacker can adapt to the augmented system.
    We show that we successfully force the attacker into the audible range, causing an average of 24.33\,dB added noise to the adversarial examples.
    We could not find adversarial examples when applying very aggressive filtering; however, this causes a drop in the benign performance. 
    
    \item \textbf{User Study.} 
    To study the auditory quality of adversarial examples, we perform a user study with an extensive crowd-sourced listening test. 
    Our results demonstrate that the adversarial examples against our system are significantly more perceptible by humans.
\end{itemize}

\smallskip
To support further research in this area, we open-source our prototype implementation, our pre-trained models, and audio samples online at \href{https://github.com/rub-syssec/dompteur}{github.com/rub-syssec/dompteur}.

%% file: sections/02_background.tex
\section{Technical Background}
\label{sec:background}

In the following, we discuss the background necessary to understand our augmentation of the \ac{ASR} system. For this purpose, we briefly introduce the fundamental concepts of \ac{ASR}s and give an overview of adversarial examples. Since our approach fundamentally relies on psychoacoustic modeling, we also explain masking effects in human perception.

\vspace{-0.5em}
\paragraph{Speech Recognition} 
\label{sec:background:asr}
\ac{ASR} constitutes the computational core of today's voice interfaces. Given an audio signal, the task of an \ac{ASR} system is to transcribe any spoken content automatically. For this purpose, traditionally, purely statistical models were used.
They now have been replaced by modern systems based on deep learning methods~\cite{bourlard-94-connectionist, hannun-14-deepspeech, graves-14-speech}, often in the form of hybrid neural/statistical models~\cite{kang-18-recurrent}.

In this paper, we consider the open-source toolkit \kaldi~\cite{povey-11-kaldi} as an example of such a modern hybrid system. Its high performance on many benchmark tasks has led to its broad use throughout the research community as well as in commercial products like \eg Amazon's Alexa~\cite{du-16-chime-4, kanda-18-chime-5, medennikov-18-chime-5}.

\kaldi, and similar DNN/HMM hybrid systems can generally be described as three-stage systems:

\begin{enumerate}
    \item \textit{Feature Extraction.}
    For the feature extraction, a frame-wise  \ac{DFT} is performed on the raw audio data to retrieve a frequency representation of the input signal. The input features of the \ac{DNN} are often given by the log-scaled magnitudes of the \ac{DFT}-transformed signal.
    \item \textit{Acoustic Model \ac{DNN}.}
    The \ac{DNN} acts as the \emph{acoustic model} of the \ac{ASR} system. It calculates the probabilities for each of the distinct speech sounds (called \emph{phones}) of its trained language being present in each time frame from its \ac{DFT} input features. Alternatively, it may compute probabilities, not of phones, but of so-called \emph{clustered tri-phones} or, more generally, of data-driven units termed \emph{senones}. 
    \item \textit{Decoding.} 
    The output matrix of the \ac{DNN} is used together with an \ac{HMM}-based language model to find the most likely sequence of words, \ie the most probable transcription. For this purpose, a dynamic programming algorithm, e.g., Viterbi decoding, is used to search the best path through the underlying \ac{HMM}. The language model describes the probabilities of word sequences, and the acoustic model output gives the probability of being in each \ac{HMM} state at each time.
\end{enumerate}

\paragraph{Psychoacoustic Modeling} 
\label{sec:psychoacoustic-modeling}
\input{includes/figure-psycho}

Recent attacks against \ac{ASR} systems exploit intrinsics of the human auditory system to make adversarial examples less conspicuous \cite{schoenherr-19-psychoacoustics, qin-19-robust, abdullah-19-practical, szurley-19-perceptual}. Specifically, these attacks utilize limitations of human perception to hide modifications of the input audio signal within inaudible ranges. We use the same effects for our approach to \emph{remove} inaudible components from the input:
\begin{itemize}
    \item \textit{Absolute Hearing Threshold.}
    Human listeners can only perceive sounds in a limited frequency range, 
    which diminishes with age. Moreover, for each frequency, the sound pressure is important to determine whether the signal component is in the audible range for humans. Measuring the \emph{hearing thresholds}, \ie the necessary sound pressures for each frequency to be audible in otherwise quiet environments, one can determine the so-called \emph{absolute hearing threshold} as depicted in Figure~\ref{fig:threshold_mask}. Generally speaking, everything above the \emph{absolute hearing thresholds} is perceptible in principle by humans, which is not the case for the area under the curve. As can be seen, much more energy is required for a signal to be perceived at the lower and higher frequencies. Note that the described thresholds only hold for cases where no other sound is present.
    \item \textit{Frequency Masking.}
     The presence of another sound---a so-called \emph{masking tone}---can change the described \emph{hearing thresholds} to cover a larger area. This \emph{masking effect} of the masking tone depends on its sound pressure and frequency. Figure~\ref{fig:frequ_mask} shows an example of a $1$\,kHz masking tone, with its induced changes of the \emph{hearing thresholds} indicated by the dashed line.
    \item \textit{Temporal Masking.}
    Like frequency masking, temporal masking is also caused by other sounds, but these sounds have the same frequency as the masked tone and are close to it in the time domain, as shown in Figure~\ref{fig:time_mask}. Its root cause lies in the fact that the auditory system needs a certain amount of time, in the range of a few hundreds of milliseconds, to recover after processing a higher-energy sound event to be able to perceive a new, less energetic sound. Interestingly, this effect does not only occur at the end of a sound but also, although much less distinct, at the beginning of a sound. This seeming causal contradiction can be explained by the processing of the sound in the human auditory system.
\end{itemize}

\paragraph{Adversarial Examples}
\label{sec:examples}

Since the seminal papers by Szegedy~\etal{}~\cite{szegedy-13-intriguing} and Biggio~\etal{}~\cite{biggio-13-evasion}, a field of research has formed around adversarial examples.
The basic idea is simple:
An attacker starts with a valid input to a machine learning system.
Then, they add small perturbations to that input with the ultimate goal of changing the resulting prediction (or in our case, the transcription of the \ac{ASR}).

More formally, given a machine learning model $f$ and an input-prediction pair $\langle\,x,\; y\,\rangle$, where $f(x) = y$, we want to find a small perturbation $\delta$ s.t.:
\begin{equation*}
    x' = x + \delta \quad \wedge \quad f(x') \neq f(x).
\end{equation*}

In this paper, we consider a stronger type of attack, a targeted one.
This has two reasons: the first is that an untargeted attack in the audio domain is fairly easy to achieve. The second is that a targeted attack provides a far more appealing (and thus, far more threatening) real-life use case for adversarial examples.
More formally, the attacker wants to perturb an input phrase $x$ (i.e., an audio signal) with a transcription~$y$ (e.g., ``Play the Beatles'') in such a way that the \ac{ASR} transcribes an attacker-chosen transcription $y'$ (e.g., ``Unlock the front door''). This can be achieved by computing an adversarial example $x'$ based on a small adversarial perturbation $\delta$ s.t.:
\begin{equation} \label{eq:1}
    x' = x + \delta \quad \wedge \quad ASR(x') = y' \quad \wedge \quad y \neq y'.
\end{equation}

There exist a multitude of techniques for creating such adversarial examples.
We use the method introduced by Sch\"onherr~\etal{}~\cite{schoenherr-19-psychoacoustics} for our evaluation in Section~\ref{sec:evaluation}.
The method can be divided into three parts: 
In a first step, attackers choose a fixed output matrix of the DNN to maximize the probability of obtaining their desired transcription $y'$.
As introduced before, this matrix is used in the \ac{ASR} system's decoding step to obtain the final transcription.
They then utilize gradient descent to perturb a starting input $x$ (i.\,e., an audio signal feed into the \ac{DNN}), to obtain a new input $x'$, which produces the desired matrix.
This approach is generally chosen in white-box attacks~\cite{yuan-18-commandersong,schoenherr-20-imperio}.
Note that we omit the feature extraction part of the \ac{ASR}; however, Sch\" onherr~\etal{} have shown that this part can be integrated into the gradient step itself~\cite{schoenherr-19-psychoacoustics}.
A third (optional) step is to utilize psychoacoustic hearing thresholds to restrict the added perturbations to inaudible frequency ranges.
More technical details can be found in the original publication~\cite{schoenherr-19-psychoacoustics}.

%% file: includes/figure-psycho.tex
\begin{figure}[t!]
  \centering
  \begin{subfigure}{\columnwidth}
  \centering
  \includegraphics[trim=0 8 0 0, clip, width=\columnwidth]{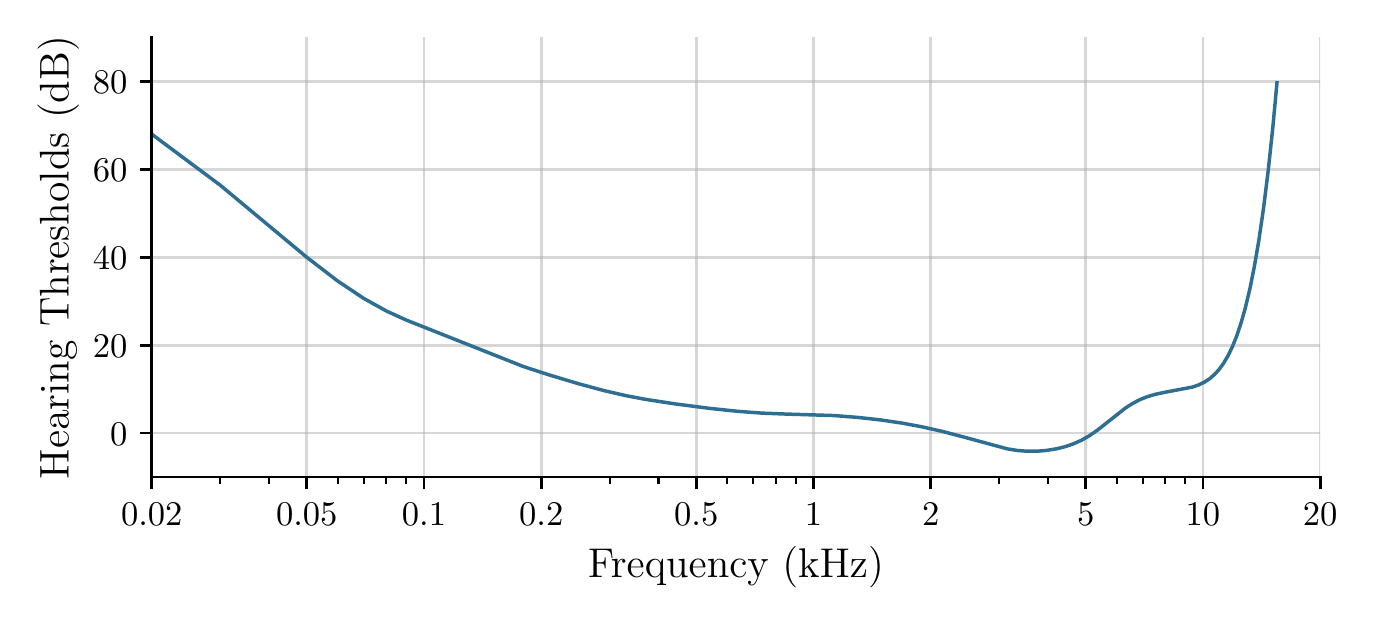}
  \caption{Absolute Hearing Thresholds}
  \label{fig:threshold_mask}
  \end{subfigure}\vspace{1em}

  \begin{subfigure}{\columnwidth}
  \centering
  \includegraphics[trim=0 8 0 0, clip, width=\columnwidth]{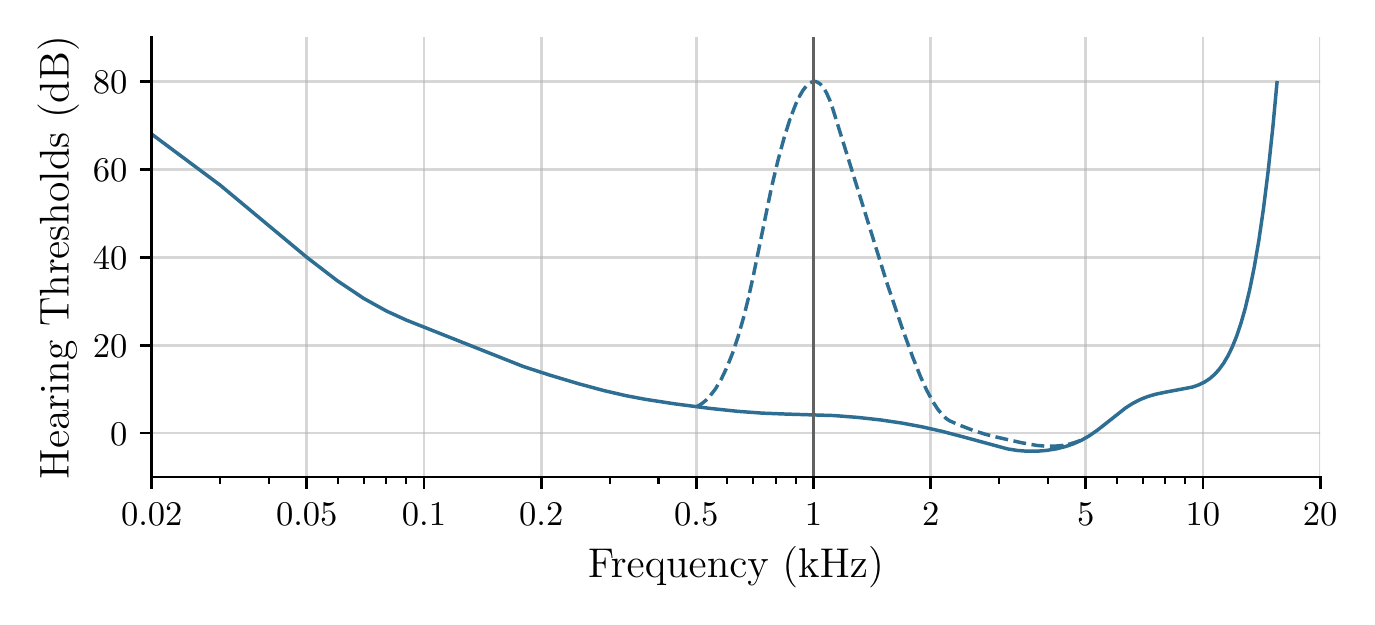}
  \caption{Frequency Masking}
  \label{fig:frequ_mask}
  \end{subfigure}\vspace{0.5em}
  
  \begin{subfigure}{\columnwidth}
  \centering
  \includegraphics[trim=0 8 0 0, clip, width=\columnwidth]{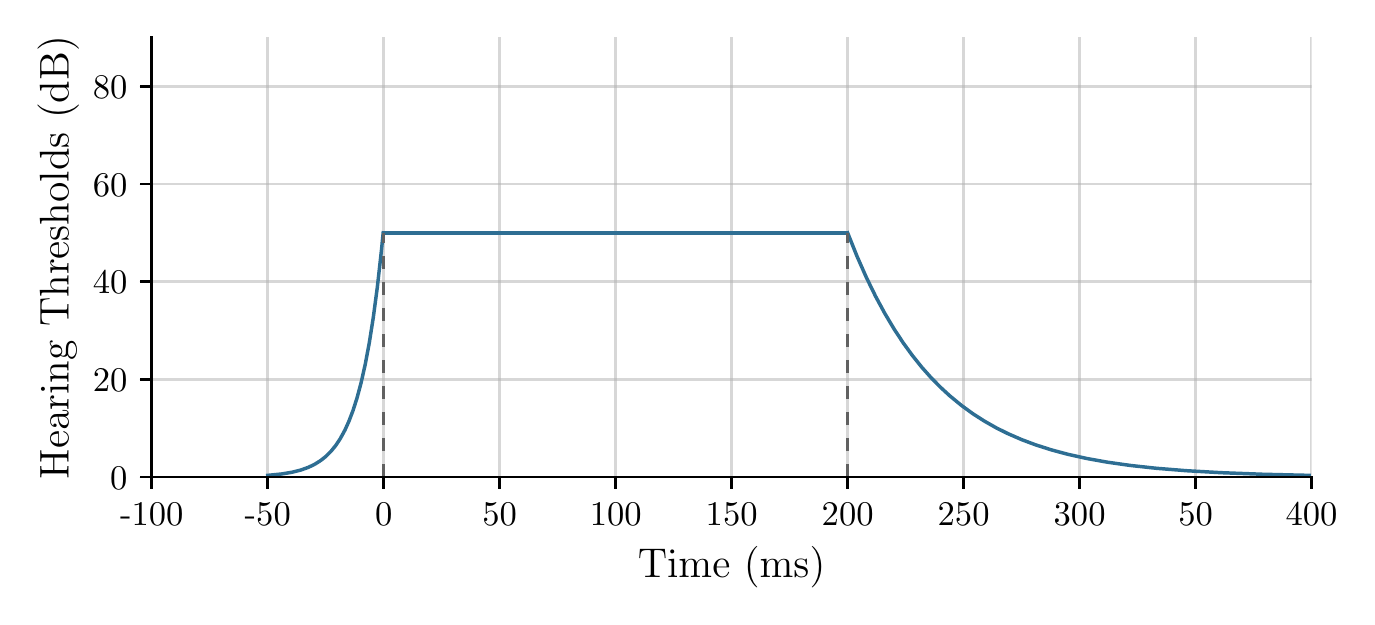}
  \caption{Temporal Masking}
  \label{fig:time_mask}
  \end{subfigure}

  \caption{\textbf{Psychoacoustic allows to describe limitations of the human auditory system.} Figure~\ref{fig:threshold_mask} shows the average human hearing threshold in quiet. Figure~\ref{fig:frequ_mask} shows an example of masking, illustrating how a loud tone at 1kHz shifts the hearing thresholds of nearby frequencies and Figure~\ref{fig:time_mask} shows how the recovery time of the auditory system after processing a loud signal leads to temporal masking.}
  \label{fig:psychoacoustics}
\end{figure}

%% file: sections/03_approach.tex
\section{Modeling the Human Auditory System}
\label{sec:approach}

We now motivate and explain our design to better align the \ac{ASR} system with human perception.
Our approach is based on the fact that the human auditory system only uses a subset of the information contained in an audio signal to form an understanding of its content.
In contrast, \ac{ASR} systems are not limited to specific input ranges and utilize every available signal -- even those \emph{inaudible} for the human auditory system.
Consequently, an attacker can easily hide changes within those ranges.
Intuitively, the smaller the overlap between these two worlds, the harder it becomes for an attacker to add malicious perturbations that are inaudible to a human listener.
This is akin to reducing the attack surface in traditional systems security.

\vspace{0.25em}
To tackle these issues, we leverage the following two design principles in our approach:
\begin{itemize}
    \item[(i)] \emph{Removing inaudible parts:}
    As discussed in Section~\ref{sec:psychoacoustic-modeling}, audio signals typically carry information imperceptible to human listeners. 
    Thus, before passing the input to the network, we utilize psychoacoustic modeling to remove these parts.
    \item[(ii)] \emph{Restricting frequency access:} 
    The human voice frequency range is limited to a band of approximately $300-5000\,Hz$~\cite{monson-14-perceptual}.
    Thus, we implement a band-pass filter between the feature extraction and model stage (\cf Section~\ref{sec:background:asr}) to restrict the acoustic model to the appropriate frequencies.
\end{itemize}

\subsection{Implementation}
\label{sec:approach:implementation}

In the following, we present an overview of the implementation of our proposed augmentations.
We extend the state-of-the-art \ac{ASR} toolkit \kaldi with our augmentations to build a prototype implementation called \tool.
Note that our proposed methods are universal and can be applied to \emph{any} \ac{ASR} system.

\paragraph{Psychoacoustic Filtering}
\label{sec:psychoacoustic-filtering}
Based on the psychoacoustic model of MPEG-1 \cite{iso-93-part3}, we use psychoacoustic hearing thresholds to remove parts of the audio that are not perceivable to humans. These thresholds define how dependencies between certain frequencies can mask, i.e., make inaudible, other parts of an audio signal.
Intuitively, these parts of the signal should not contribute any information to the recognizer.
They do, however, provide space for an attacker to hide adversarial noise.

We compare the absolute values of the complex valued \ac{STFT} representation of the audio signal~$\complexsignal$ with the hearing thresholds~$\hearingthresholds$ and define a mask via

\begin{equation}
\mask(\frameindex, \freqindex) = 
      \begin{cases}
    0 & \quad \text{if } \complexsignal(\frameindex,\freqindex) \leq \hearingthresholds(\frameindex,\freqindex) + \margin \\
    1 & \quad \text{else}
  \end{cases},
  \label{eq:hearing}
\end{equation}

with $\frameindex = 0, \dots ,\framemax - 1 $ and $ \freqindex = 0, \dots ,\frequmax - 1$.
We use the parameter~$\margin$ to control the effect of the hearing thresholds.
For $\margin=0$, we use the original hearing threshold, for higher values we use a more aggressive filtering, and for smaller values we retain more from the original signal.
We explore this in detail in Section~\ref{sec:evaluation}.
We then multiply all values of the signal~$\complexsignal$ with the mask~$\mask$
\begin{equation}
    \inputsignal = \complexsignal \odot \mask, 
    \label{eq:psycho}
\end{equation}
to obtain the filtered signal $\inputsignal$.

\paragraph{Band-Pass Filter}

High and low frequencies are not part of human speech and do not contribute significant information.
Yet, they can again provide space for an attacker to hide adversarial noise.
For this reason, we remove low and high frequencies of the audio signal in the frequency domain.
We apply a band-pass filter after the feature extraction of the system by discarding those frequencies that are smaller or larger than certain thresholds (the so-called cut-off frequencies).
Formally, the filtering can be described via

\begin{equation}
\inputsignal(\frameindex,\freqindex) = 0 \quad \forall \, f_{\text{max}} < \freqindex < f_{\text{min}},
\end{equation}

where $f_{\text{max}}$ and $f_{\text{min}}$ describe the lower and the upper cut-off frequencies of the band-pass.

\subsection{Attacker Model}
\label{sec:approach:attacker}

While some of our augmentations improve the \ac{ASR} system's overall performance, we are specifically interested in its performance against adversarial perturbations.
To achieve any meaningful results, we believe the attacker needs to have \emph{complete} control over the input.
Following guidelines recently established by Carlini~\etal{}~\cite{carlini-19-evaluating}, we embark from theoretical attack vectors towards the definition of a realistic threat model, capturing real-world capabilities of attackers.

The key underlying insight is that the amount of perturbations caused by a real-world attack cannot be limited.
This is easy to see: in the worst case, the attacker can always force the target output by replacing the input with the corresponding audio command. 
Note that this, in turn, implies that we cannot completely prevent adversarial attacks \emph{without} also restricting benign inputs.

We can also not rely on obfuscation.
Previous works have successfully shown so-called parameter-stealing attacks, which 
build an approximation of a black-box system~\cite{ilyas-17-blackbox, papernot-16-transferability, tramer-16-stealing, papernot-17-blackbox, wang-18-stealing}. 
Since an attacker has full control over this approximated model, they can utilize powerful white-box attacks against it, which transfer to the black-box model.\smallskip

In summary, we use the following attacker model:
\begin{itemize}
    \item \emph{Attacker Knowledge:} Following Kerckhoffs' principle~\cite{kerckhoffs-83-cryptographic}, we consider a \emph{white-box} scenario, where the attacker has complete knowledge of the system, including all model parameters, training data, etc.

    \item\emph{Attacker Goals:} To maximize practical impact, we assume a targeted attack, \ie the attacker attempts to perturb a given input $x$ to fool a speech recognition system into outputting a false, \emph{attacker-controlled} target transcription $y'$ based on Equation~\eqref{eq:1}.
    
    \item \emph{Attacker Capabilities:} The attacker is granted complete control over the input, and we explicitly do not restrict them in any way on how $\delta$ should be crafted. Note, however, that $\delta$ is commonly minimized during computation according to some distance metric. For example, by measuring the \emph{perceived} noise, an attacker might try to minimize the conspicuousness of their attack~\cite{schoenherr-19-psychoacoustics}.
    
\end{itemize}

We choose this attacker model with the following in mind: We aim to limit the attacker, not in the amount of applied perturbations, but rather confine the nature of perturbations themselves. 
In particular, we want adversarial perturbations to be clearly perceptible by humans and, thus, strongly perturb the initial input such that the added noise becomes audible for a human listener. In this case, an attack---although still viable---significantly loses its malicious impact in practice.

%% file: sections/04_results.tex
\section{Evaluation}
\label{sec:evaluation}

With the help of the following experiments, we empirically verify and assess our proposed approach according to the following three main aspects:

\begin{itemize}
    \item[(i)] \emph{Benign Performance.} The augmentation of the system should impair the performance on benign input as little as possible. 
    We analyze different parameter combinations for the psychoacoustics and our band-pass filter to show that our augmented model retains its practical use.

    \item[(ii)] \emph{Adaptive Attacker.} To analyze the efficacy of the augmented system, we construct and assess its robustness against adversarial examples generated by a strong attacker with white-box access to the system. This attacker is aware of our augmentations and \emph{actively} factors them into the optimization.
    
    \item[(iii)] \emph{Listening Test.} Finally, we verify the success of our method by a crowd-sourced user study.
    We conduct a listening test, investigating the quality (i.e., the inconspicuousness) of the adversarial examples computed from the adaptive attacker against the augmented \ac{ASR} system.
\end{itemize}

All experiments were performed on a server running Ubuntu 18.04, with 128 GB RAM, an Intel Xeon Gold 6130 CPU, and four Nvidia GeForce RTX 2080 Ti.
For our experiments, we use \kaldi in version 5.3 and train the system with the default settings from the \ac{WSJ} training recipe.

\subsection{Metrics}
To assess the quality of adversarial examples both in terms of efficacy and inconspicuousness, we use two standard measures.

\paragraph{Word Error Rate (WER)}
The \ac{WER} is computed based on the Levenshtein distance~\cite{navarro-01-levenshtein}, which describes the \emph{edit distance} between the reference transcription and the ASR output (i.e., the minimum number of edits required to transform the output text of the \ac{ASR} system into the correct text).

We compute the Levenshtein distance $\mathcal{L}$ as the sum over all substituted words $S$, inserted words $I$, and deleted words~$D$:

$$\text{WER} = 100 \cdot \frac{\mathcal{L}}{N} = 100 \cdot \frac{S + D + I}{N},$$ 
where $N$ is the total number of words of the reference text.
The smaller the \ac{WER}, the fewer errors were made by the \ac{ASR} system.

To evaluate the efficacy of adversarial examples, we measure the \ac{WER} between the adversarial target transcription and the output of the \ac{ASR} system. Thus, a \emph{successful adversarial example} has a \ac{WER} of 0\,\%, \ie fully matching the desired target description~$y'$. 
Note that the \ac{WER} can also reach values above 100\,\,\%, e.\,g., when many words are inserted. This can especially happen with unsuccessful adversarial examples, where mostly the original text is transcribed, which leads to many insertions. 

\paragraph{Segmental Signal-to-Noise Ratio (SNRseg)}
The \ac{WER} can only measure the success of an adversarial example in fooling an \ac{ASR} system.
For a real attack, we are also interested in the (in-)\,conspicuousness of adversarial examples, \ie the level of the added perturbations. 
For this purpose, we quantify the changes that an attacker applies to the audio signal.
Specifically, we use the \ac{SNR} to measure the added perturbations.
More precisely, we compute the \ac{SNRseg} \cite{voran-1995-perception, yang-99-enhancedmb}, a more accurate measure of distortion than the \ac{SNR}, when signals are aligned \cite{yang-99-enhancedmb}. 

Given the original audio signal $x(t)$ and the adversarial perturbations $\sigma(t)$ defined over the sample index $t$, the \ac{SNRseg} can be computed via
$$\text{SNRseg(dB)} = \dfrac{10}{K} \sum_{k=0}^{K-1} \log _{10}  \frac{\sum_{t=Tk}^{Tk+T-1} x^2(t)}{\sum_{t=Tk}^{Tk+T-1} \sigma^2 (t)},$$
with $T$ being the number of samples in a segment and $K$ the total number of segments. For our experiments, we set the segment length to 16\,ms, which corresponds to $T = 256$ samples for a 16\,kHz sampling rate.

The \emph{higher} the \ac{SNRseg}, the \emph{less} noise has been added to the audio signal.
Hence, an adversarial example is considered less conspicuous for higher \ac{SNRseg} values.
Note that we use the \ac{SNRseg} ratio only as an approximation for the perceived noise. We perform a listening test with humans for a realistic assessment and show that the results of the listening test correlate with the reported \ac{SNRseg} (cf. Section~\ref{sec:user-study}). 


\input{includes/figure-benign-accuracy-bandpass-heatmap}

\subsection{Benign Performance}
\label{sec:benign-performance}
Our goal is to preserve accuracy on benign inputs (\ie non-malicious, unaltered speech) while simultaneously impeding an attacker as much as possible.
To retain that accuracy as much as possible, the parameters of the band-pass, and the psychoacoustic filter need to be carefully adjusted.
Note that adversarial robustness is generally correlated with a loss in accuracy for image classification models~\cite{tsipras-19-robustness}.
Accordingly, we assume that higher adversarial robustness likely incurs a trade-off on benign input performance.

All models in this section are trained with the default settings for the \emph{Wall Street Journal} (WSJ) training recipe of the \kaldi toolkit \cite{povey-11-kaldi}.  
The corresponding \ac{WSJ}-based speech corpus \cite{paul-92-wsj} contains approximately 81 hours of training data and consists of uttered sentences from the Wall Street Journal.

We train three models for each configuration and report the WER on the test set for the model with the best performance. For the test set, we use the \texttt{eval92} subset consisting of 333 utterances with a combined length of approximately 42 minutes.

\paragraph{Band-Pass Filtering} The band-pass filter limits the signal's frequency range by removing frequencies below and above certain thresholds.
Our goal is to remove parts of the audio that are not used by the human voice.
We treat these values as classical hyperparameters and select the best performing combination by grid searching over different cut-off frequencies; for each combination, we train a model from scratch, using the training procedure outlined above. 
The results are depicted in Figure~\ref{fig:benign-accuracy-bandpass-heatmap}.
If we narrow the filtered band (\ie remove more information), the WER gradually increases and, therefore, worsens the recognizer's accuracy. 
However, for many cases, even when removing a significant fraction of the signal, the augmented system either achieves comparable results or even surpasses the baseline (WER $5.90\,\%$).
In the best case, we reach an improvement by $0.35\,\%$ percentage points to a WER of $5.55\,\%$ (200\,Hz-7000\,Hz). 
This serves as evidence that the unmodified input contains signals that are not needed for transcription.
In Section~\ref{sec:narrower-band}, we further confirm this insight by analyzing models with narrower bands.
We hypothesize that incorporating a band-pass filter might generally improve the performance of \ac{ASR} systems but note that further research on this is needed.

For the remaining experiments, if not indicated otherwise, we use the 200-7000\,Hz band-pass.

\paragraph{Psychoacoustic Filtering} The band-pass filter allows us to remove high- and low-frequency parts of the signal; however, the attacker can still hide within this band in inaudible ranges.
Therefore, we use psychoacoustic filtering as described in Section~\ref{sec:psychoacoustic-filtering} to remove these parts in the signal. 
We evaluate different settings for $\margin$ from Equation~\eqref{eq:hearing} -- by increasing $\margin$, we artificially increase the hearing thresholds, resulting in more aggressive filtering. 
We plot the results in Figure~\ref{fig:benign-accuracy-phi} for both psychoacoustic filtering and a baseline WER, with and without band-pass, respectively. 
The WER increases with increasing $\margin$, \ie the performance drops if more of the signal is removed, independent of the band-pass filter.

\input{includes/table-benign-accuracy}
\input{includes/figure-benign-accuracy-phi}

When we use no band-pass filter, the WER increases from $5.90\,\%$ (baseline) to $6.50\,\%$ for $\margin = 0$\,dB, which is equivalent to removing everything below the human hearing thresholds.
When we use more aggressive filtering---which results in better adversarial robustness (\cf Section~\ref{sec:adaptive_attack})---the WER increases up to $8.05\,\%$ for $\margin = 14$\,dB.
Note that the benefits of the band-pass filter remain even in the presence of psychoacoustic filtering and results in improving the WER to $6.10\,\,\%$ ($\margin = 0$\,dB) and $7.83\,\,\%$ ($\margin = 14$\,dB).
We take this as another sign that a band-pass filter might generally be applicable to \ac{ASR} systems.

\paragraph{Cross-Model Benign Accuracy} Finally, we want to evaluate if \tool indeed only uses relevant information.
To test this hypothesis, we compare three different models.
One completely unaugmented model (\ie an unmodified version of \kaldi), one model trained with psychoacoustics filtering, and one model trained with both psychoacoustics filtering and a band-pass filter.
We feed these models two types of inputs: (i) \emph{standard inputs}, \ie inputs directly lifted from the WSJ training set, and (ii) \emph{processed inputs}, these inputs are processed by our psychoacoustic filtering.
If our intuitive understanding is correct and \tool does indeed learn a better model of the human auditory system, it should retain a low WER even when presented with non-filtered input.
Thus, the model has learned to \emph{ignore} unnecessary parts of the input.
The results are shown in Table~\ref{table:benign-accuracy} and match our hypothesis:
\tool{}'s performance only drops slightly ($6.10\,\% \to 6.33\,\%$) when presented with unfiltered input or does even improve if the band-pass is disabled ($6.50\,\% \to 6.20\,\%$).
\kaldi, on the other hand, heavily relies on this information when transcribing audio, increasing its WER by $2.84$ percentage point ($5.90\,\% \to 8.74\,\%$).
Thus, the results further substantiate our intuition that we filter only irrelevant information with our~approach.

\input{includes/figure-attack-itr}

\subsection{Adaptive Attacker}
\label{sec:adaptive_attack}
\input{includes/table-adaptive-attack}

We now want to evaluate how robust \tool is against adversarial examples.
We construct a strong attacker with complete knowledge about the system and, in particular, our modifications. Ultimately, this allows us to create successfully adversarial examples.
However, as inaudible ranges are removed, the attacker is now forced into human-perceptible ranges, and, consequently, the attack loses much of its malicious impact.
We provide further support for this claim in Section~\ref{sec:user-study} by performing a user study to measure the perceived quality of adversarial examples computed with this~attack.

\paragraph{Attack.}
We base our evaluation on the attack by~Sch{\"o}nherr~\etal~\cite{schoenherr-19-psychoacoustics}, which presented a strong attack that works with~\kaldi. 
Recent results show that it is crucial to design adaptive attacks as simple as possible while simultaneously resolving any obstacles for the optimization \cite{tramer-20-adaptiveattacks}.
To design such an attacker against \tool, we need to adjust the attack to consider the augmentations in the optimization.
Therefore, we extend the baseline attack against \kaldi to include both the band-pass and psychoacoustic filter into the computation.
This allows the attacker to compute gradients for the entire model in a white-box fashion.

More specifically, we extend the gradient descent step s.t.~(i) the band-pass filter and (ii) the psychoacoustic filter component back-propagates the gradient respectively. 
\begin{itemize}
    \item[(i)] \emph{Band-Pass Filter.} For the band-pass filter we remove those frequencies that are smaller and larger than the cut-off frequencies of the band-pass filter. This is also applied to the gradients of the back propagated gradient to ignore changes that will fall into ranges of the removed signal
    \begin{equation}
        \nabla_{\inputsignal(\frameindex,\freqindex)} = 0 \quad \forall f_{\text{max}} < \freqindex < f_{\text{min}}.
    \end{equation}
    \item[(ii)] \emph{Psychoacoustic Filter.} The same principle is used for the psychoacoustic filtering, where we use the mask $\mask$ to zero out components of the signal that the network will not process
    \begin{equation}
        \nabla_{\complexsignal} = \nabla_{\inputsignal} \odot \mask.
    \end{equation}
\end{itemize}

\input{includes/figure-spectograms}

\paragraph{Experimental Setup.}
We evaluate the attack against different versions of \tool.
Each model uses a $200-7000\,Hz$ band-pass filter, and we vary the degrees of the psychoacoustic filtering ($\margin \in \{0, 3, 6, 9, 12, 13, 14\}$).
We compare the results against two baselines to evaluate the inconspicuousness of the created adversarial examples.
First, we run the attack of Sch{\"o}nherr~\etal without psychoacoustic hiding against an unaltered version \kaldi.
Second, we re-enable psychoacoustic hiding and run the original attack against \kaldi, to generate state-of-the-art inaudible adversarial examples.
As a sanity check, we also ran the original attack (\ie with psychoacoustic hiding) against \tool. 
As expected, this attack did not create any adversarial examples since we filter the explicit ranges the attacker wants to utilize.

As a target for all configurations, we select 50 utterances with an approximate length of 5s from the WSJ speech corpus test set \texttt{eval92}. The exact subset can be found in appendix \ref{app:targets}. We use the same target sentence \emph{send secret financial report} for all samples. 

These parameters are chosen such that an attacker needs to introduce \textapprox$4.8$ phones per second into the target audio, which Sch{\"o}nherr~\etal suggests as both effective and efficiently possible \cite{schoenherr-19-psychoacoustics}. Furthermore, we picked the utterances and target sentence to be \emph{easy} for an attacker in order to decouple the influence on our analysis. Specifically, for these targets the baseline has a very high success rate and low SNRseg (cf. Table \ref{table:adaptive-attack}). Note that the attack is capable of introducing arbitrary target sentences (up to a certain length).
In Section \ref{sec:target-phone-rate}, we further analyze the influence of the phone~rate, and in particular, the influence of the target utterance and sentence on the SNRseg.
We compute adversarial examples for different learning rates and a maximum of 2000 iterations. This number is sufficient for the attack to converge, as shown in Figure \ref{fig:attack-itr}, where the WER is plotted as a function of the number of iterations.

\paragraph{Results.}
The main results are summarized in Table~\ref{table:adaptive-attack}.
We report the average SNRseg over all adversarial examples, the best ($\text{SNRseg}_\text{max}$), and the number of successful adversarial examples created.

We evaluate the attack using different learning rates ($0.05$, $0.10$, $0.5$, and $1$).
In our experiments, we observed that while small learning rates generally produce less noisy adversarial examples, they simultaneously get more stuck in local optima.
Thus, to simulate an attacker that would run an extensive search and uses the best result we also report the intersection of successful adversarial examples over all learning rates. If success rate is the primary goal, we recommend a higher learning rate.

By increasing $\margin$, we can successfully force the attacker into audible ranges while also decreasing the attack's success rate.
When using very aggressive filtering ($\margin=14$), we can prevent the creation of adversarial examples completely, albeit with a hit on the benign WER ($5.55\,\% \to 7.83\,\%$).
Note, however, that we only examined 50 samples of the test corpus, and other samples might still produce valid adversarial examples.
We see that adversarial examples for the augmented systems are more distorted for all configurations compared to the baselines. 
When using $\margin \geq 12$, we force a negative SNRseg for all learning rates. For these adversarial examples, the noise (\ie adversarial perturbations) energy exceeds the energy of the signal.
With respect to the baselines, the noise energy increases on average by 21.42\,dB (without psychoacoustic hiding) and 24.33\,dB (with hiding enabled).
This means there is, on average, ten times more energy in the adversarial perturbations than in the original audio signal.
A graphical illustration can be found in Figure~\ref{fig:spectogram}, where we plot the power spectra of different adversarial examples compared to the original signal.

\subsubsection{Non-speech Audio Content}
\input{includes/table-audio-content}

The task of an ASR system is to transcribe audio files with spoken content.
An attacker, however, might pick other content, i.e., music or ambient noise, to obfuscate his hidden commands.
Thus, we additionally evaluated adversarial examples based on audio files containing music and bird sounds.
The results are presented in Table~\ref{table:audio-content}.

We can repeat our observations from the previous experiment.
When we utilize a more aggressive filter, we observe that the perturbation energy of adversarial examples increases with respect to the baselines by up to 24.08 dB (birds) and 21.32 dB (music).
Equally, the attack's general success decreases to 5/50 (birds) and 3/50 (music) successful adversarial examples.

Note that the SNRseg for music samples are in general higher than that of speech and bird files as these samples have a more dynamic range of signal energy.
Hence, potentially added adversarial perturbations have a smaller impact on the calculation of the SNRseg.
The absolute amount of added perturbations, however, is similar to that of other content.
Thus, when listening to the created adversarial examples\footnote{~\href{https://rub-syssec.github.io/dompteur/}{rub-syssec.github.io/dompteur}} the samples are similarly distorted. 
This is further confirmed in Section~\ref{sec:user-study} with our listening test.

\subsubsection{Target Phone Rate}
\label{sec:target-phone-rate}
\input{includes/figure-phone-rate}
\input{includes/table-bandpass}

The success of the attack depends on the ratio between the length of the audio file and the length of the target text, which we refer to as the \emph{target phone rate}.
This rate describes how many phones an attacker can hide within one second of audio content.

In our experiments, we used the default ratios recommended by Sch{\"o}nherr \etal
However, a better rate might exist for our setting.
Therefore, to evaluate the effect of the target phone rate, we sample target texts of varying lengths from the WSJ corpus and compute adversarial examples for different target phone rates.
We pick phone rates ranging from 1 to 20 and run 20 attacks for each of them for at most 1000 iterations, resulting in 400 attacks. 

The results in Figure~\ref{fig:phone-rates} show that, in general, with increasing phone rates, the SNRseg decreases and stagnates for target phone rate beyond 12. This is expected as the attacker tries to hide more phones and, consequently, needs to change the signal more drastically.
Thus, we conclude that the default settings are adequate for our setting.

\subsubsection{Band-Pass Cut-off Frequencies}
\label{sec:narrower-band}
So far, we only considered a relatively wide band-pass filter (200-7000\,Hz).
We also want to investigate other cut-off frequencies.
Thus, we disable the psychoacoustic filtering and compute adversarial examples for different models examined in Section~\ref{sec:benign-performance}.
We run the attack for each band-pass model with 20 speech samples for at most 1000 iterations. 

The results are reported in Table~\ref{table:bandpass}.
We observe that the energy amount of adversarial perturbation remains relatively constant for different filters, which is expected since the attacker has complete knowledge of the system. As we narrow the frequency band, the attacker adopts and puts more perturbation within these bands.

Apart from the SNRseg, we also observe a decrease in the attack success, especially for small high cut-off frequencies, with only 11/20 (300-3000\,Hz) and 12/20  (500-3000\,Hz) successful adversarial examples.

%% file: includes/figure-benign-accuracy-bandpass-heatmap.tex
\begin{figure}[t]
    \centering
  	\includegraphics[trim=10 15 0 15, clip, width=\columnwidth]{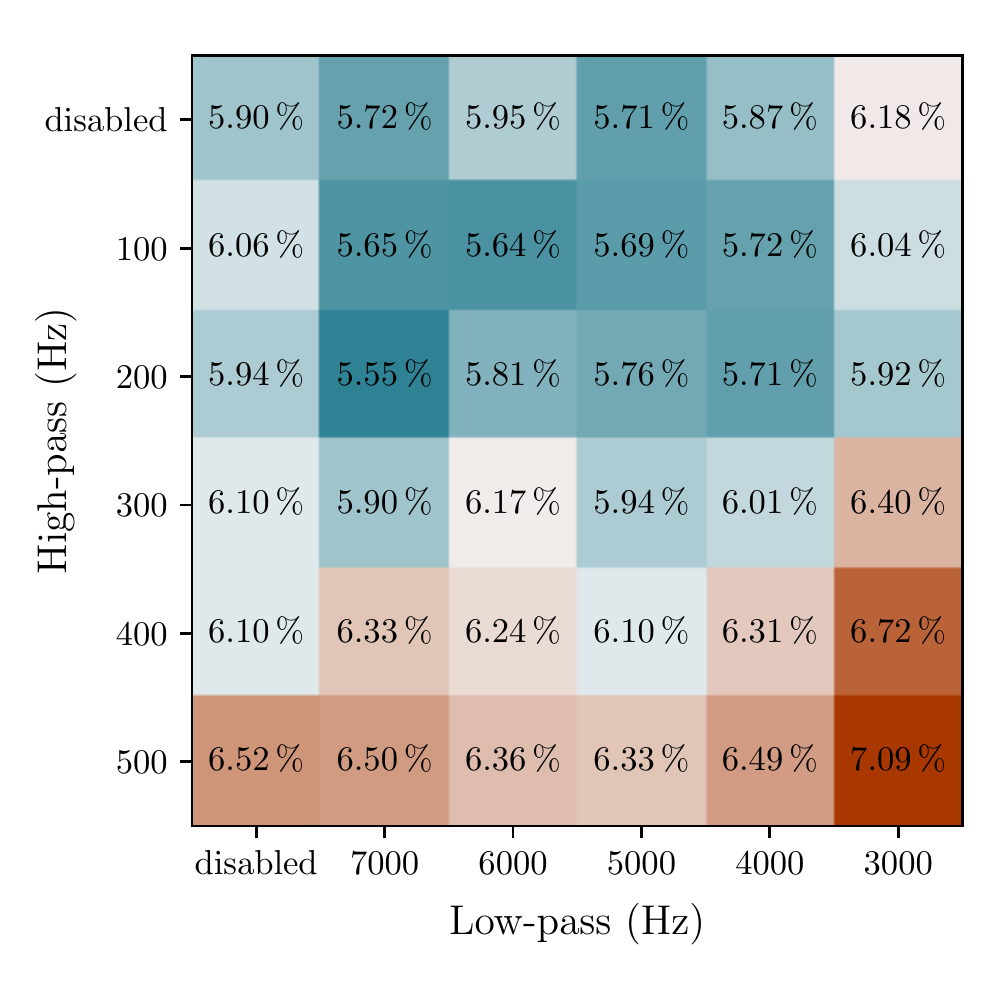}
    \caption{\textbf{Word Error Rate (WER) for different band-pass filters.} For each filter, we train three models and report the best accuracy in terms of WER (the lower, the better).}
    \label{fig:benign-accuracy-bandpass-heatmap}
\end{figure}

%% file: includes/table-benign-accuracy.tex
\begin{table}[t]
\centering
\footnotesize
\caption{\textbf{Recognition rate of the \ac{ASR} system on benign input.} We report the performance of an unmodified \kaldi system as well as two variants hardened by our approach. For our model, the scaling factor $\phi$ is set to 0 and the band-pass filter configured with 200-7000Hz.
Note, when feeding standard input to \tool, we disable its psychoacoustic filtering capabilities.
}
\resizebox{1.0\columnwidth}{!}{
\begin{tabular}{@{}lrrrrr@{}}
\toprule
                    & \quad &   \multirow{2}{4em}{\centering \kaldi \phantom{123456}} & \quad & \multicolumn{2}{c}{\tool}  \\ \cmidrule{3-3} \cmidrule{5-6}
                    &&     && \multicolumn{1}{c}{w/o band-pass}      & \multicolumn{1}{c}{w/ band-pass}       \\
\midrule
Standard\phantom{e} Input    && WER 5.90\,\%  && WER 6.20\,\%  & WER 6.33\,\%  \\
\rule{0pt}{2ex}
Processed Input     && WER 8.74\,\%  && WER 6.50\,\%  & WER 6.10\,\%  \\
\bottomrule
\end{tabular}}
\label{table:benign-accuracy}
\end{table}

%% file: includes/figure-benign-accuracy-phi.tex
\begin{figure}[t]
    \centering
  	\includegraphics[trim=10 10 0 10, clip, width=\columnwidth]{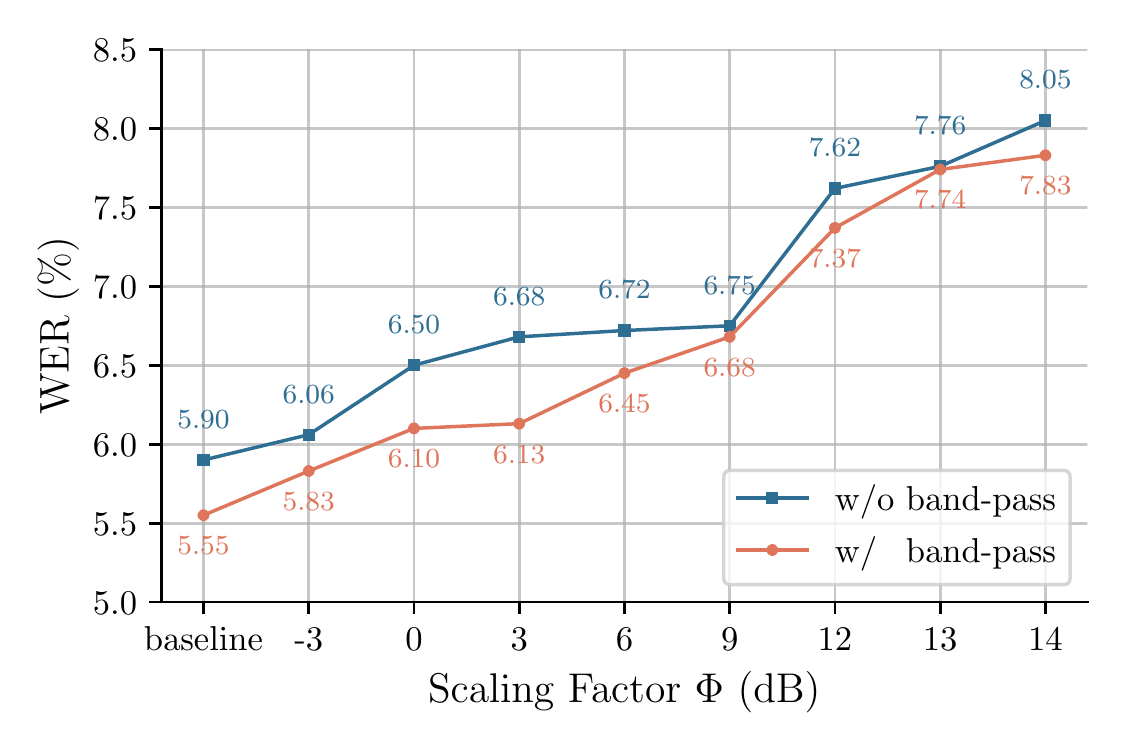}
    \caption{\textbf{Recognition rate for psychoacoustic filtering.} For each $\phi$ we train a model both with and without band-pass filter (200-7000Hz) and report the best accuracy from three repetitions. 
    A negative scaling factor partially retains inaudible ranges.
    Note that the benefits of the band-pass filter are retrained, even when we incorporate psychoacoustic filtering.
    }
    \label{fig:benign-accuracy-phi}
\end{figure}

%% file: includes/figure-attack-itr.tex
\begin{figure}[t]
    \centering
  	\includegraphics[trim=10 10 0 10, clip, width=\columnwidth]{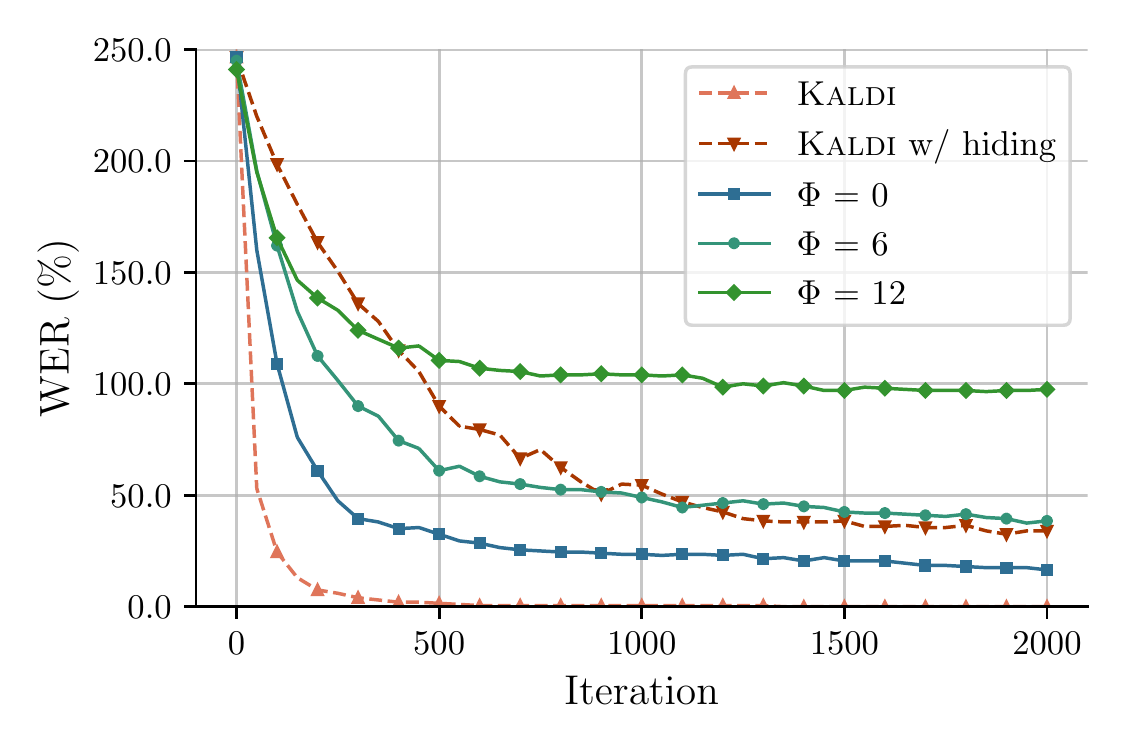}
    \caption{\textbf{Progress of attack for computing adversarial examples.} 
    We run the attack against multiple instances of \tool with different values of $\Phi$ and a 200Hz-7000Hz band-pass filter.
    The baseline refers to the attack from Schönherr et al.~\cite{schoenherr-19-psychoacoustics} against an unaltered instance of \kaldi.
    For each attack report the Word Error Rate (WER) for the first 2000 iterations.}
    \label{fig:attack-itr}
\end{figure}

%% file: includes/table-adaptive-attack.tex
\begin{table*}[t]
\centering
\footnotesize
\caption{\textbf{Number of successful Adversarial Examples (AEs) and Segmental Signal-to-Noise (SNRseg) ratio for the experiments with the adaptive attacker.} We report the numbers for all computed adversarial examples against the augmented models as well as our two baselines (with and without psychoacoustic hiding). As the success rate and SNRseg depend on the learning rate, we combine these in the last row. For this, we select the best (i.e., least noisy) AE for each utterance among the four learning rates. For the SNRseg, we only consider successful AEs.
The higher the SNSseg, the \emph{less} noise (\ie adversarial perturbation) is present in the audio signal.
Negative values indicate that the energy of the noise exceeds the energy in the original signal.
}
\astretch{1.05}
\resizebox{1.0\textwidth}{!}{
\begin{tabular}{@{}llccccccccccc@{}}
\toprule
  \multicolumn{2}{c}{} 
            & \quad & \multicolumn{2}{c}{\kaldi} 
                            & \quad & \multicolumn{7}{c}{\tool} \\ 
\cmidrule{4-5} \cmidrule{7-13}
  \multirow{2}{4em}{Learning Rate} 
& \multirow{2}{1.5em}{Metric \phantom{abcde}}  
            & \quad  & \multirow{2}{4.4em}{\centering baseline w/o hiding}
                     & \multirow{2}{4.4em}{\centering baseline w/~ hiding}  
                            & \quad &  \multirow{2}{4em}{\centering $\margin = 0$}       
                                    &  \multirow{2}{4em}{\centering $\margin = 3$} 
                                    &  \multirow{2}{4em}{\centering $\margin = 6$} 
                                    &  \multirow{2}{4em}{\centering $\margin = 9$}
                                    &  \multirow{2}{4em}{\centering $\margin = 12$}
                                    &  \multirow{2}{4em}{\centering $\margin = 13$} 
                                    &  \multirow{2}{4em}{\centering $\margin = 14$} \\\\ 
\midrule
\multirow{2}{*}{0.05}
& $AEs$   & \quad & 50/50         & 17/50         & \quad & 31/50         & 28/50         & 10/50         & ~4/50         & ~0/50         & ~0/50         & ~0/50 \\
& $SNR$   & \quad & ~~5.80/~14.44    & ~13.48/~18.50 & \quad & ~6.03/10.63   & ~~3.61/~8.31  & ~~1.21/5.53   & ~~1.50/~3.23  & ---           & ---           & ---   \\
\rule{0pt}{4ex}
\multirow{2}{*}{0.01}
& $AEs$   & \quad & 50/50         & 28/50         & \quad & 38/50         & 34/50         & 22/50         & 10/50         & ~0/50         &  ~0/50        & ~0/50 \\
& $SNR$   & \quad & ~~2.15/~10.59   & ~~9.36/~15.81 & \quad & ~3.74/~9.53   & ~~0.47/~6.41  & ~-0.68/3.60   & ~-1.31/~1.10  & ---           & ---           & ---   \\
\rule{0pt}{4ex}
\multirow{2}{*}{0.5~}
& $AEs$   & \quad & 49/50         & 23/50         & \quad & 48/50         & 44/50         &   42/50       & 20/50         & ~1/50         & ~1/50         & ~0/50 \\
& $SNR$   & \quad & ~-8.54/~-0.02   & ~~1.08/~~8.63 & \quad & -3.78/~3.24   & ~-6.51/~0.11  & ~-7.74/-1.47  & ~-8.69/-3.35  & -13.56/-13.56 & -15.69/-15.69 & ---   \\
\rule{0pt}{4ex}
\multirow{2}{*}{1~~~}
& $AEs$   & \quad & 50/50         & 16/50         & \quad & 49/50         & 50/50         & 43/50         & 23/50         & ~1/50         & ~1/50         & ~0/50 \\
& $SNR$   & \quad & -13.68/~-5.03   & ~-1.81/~~4.50 & \quad & -7.44/-0.29   & -10.50/-3.00  & -10.99/-4.34  & -11.98/-6.37  & -17.69/-17.69 & -11.73/-11.73 & ---   \\ 
\rule{0pt}{4ex}
\multirow{2}{*}{\textbf{Best AEs}} 
& \textbf{\textit{AEs}}   & \quad & \textbf{50/50}         & \textbf{37/50}         & \quad &\textbf{ 50/50}         & \textbf{50/50}         &\textbf{ 46/50}         & \textbf{27/50  }       & \textbf{~2/50 }        & \textbf{~2/50 }        & \textbf{~0/50} \\
& \textbf{\textit{SNR}}  & \quad & \textbf{~~5.80/~14.44}   & \textbf{~~8.71/~18.50} & \quad & \textbf{~3.36/10.63}   & ~\textbf{~0.85/~8.31}  & \textbf{~-4.71/5.53 }  & \textbf{~-7.14/~3.23}  & \textbf{-15.62/-13.56} & \textbf{-13.71/-11.73} & \textbf{--- }\\
\bottomrule
\end{tabular}
}
\begin{flushleft}
\vspace{-0.5em}
\scriptsize{$AEs$: Successful adversarial examples;\, $SNR$: SNRseg/$\text{SNRseg}_\text{{max}}$ in dB}\\
\end{flushleft}
\label{table:adaptive-attack}
\end{table*}

%% file: includes/figure-spectograms.tex
\begin{figure*}[t]
  \centering
  \begin{subfigure}{0.49\textwidth}
  \centering
  \includegraphics[trim=0 5 0 0, clip, width=\textwidth]{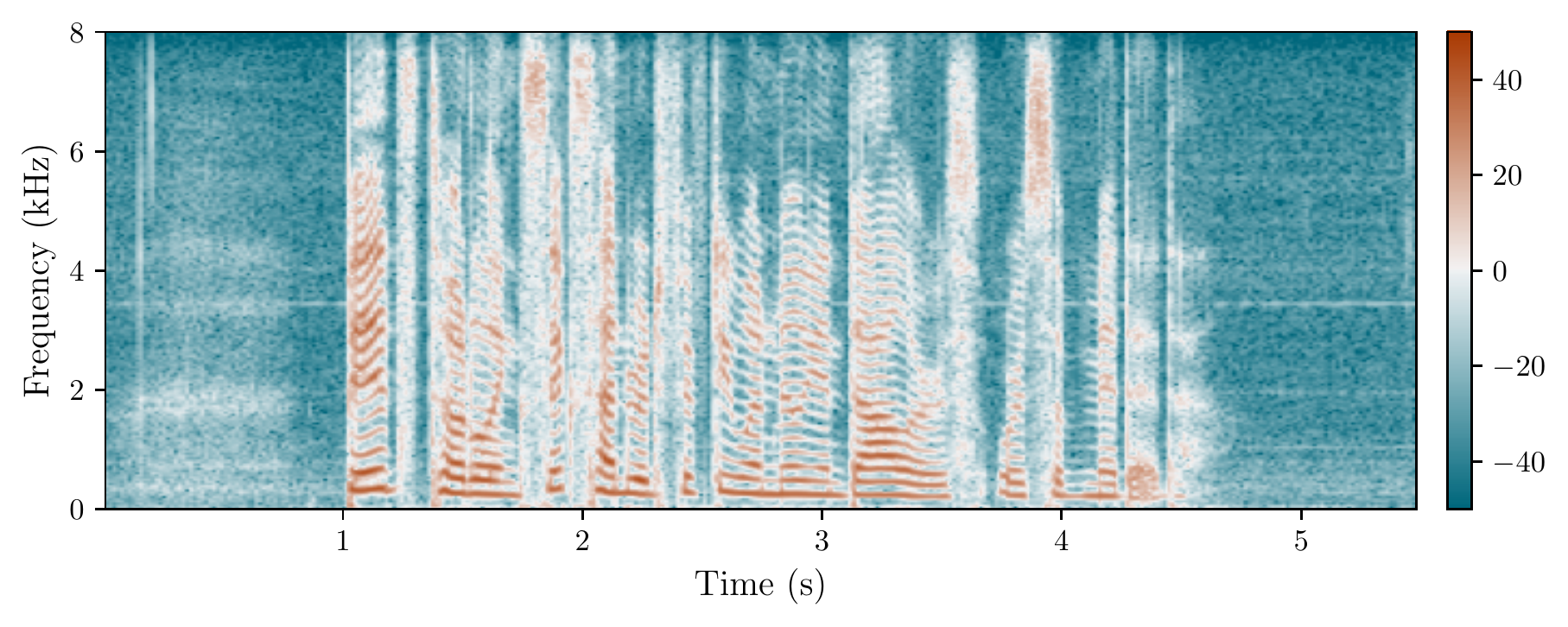}
  \caption{Unmodified Signal}
  \label{fig:spectogram-reference}
  \end{subfigure}\vspace{1em}
  \hfill
  \begin{subfigure}{0.49\textwidth}
  \centering
  \includegraphics[trim=0 5 0 0, clip, width=\textwidth]{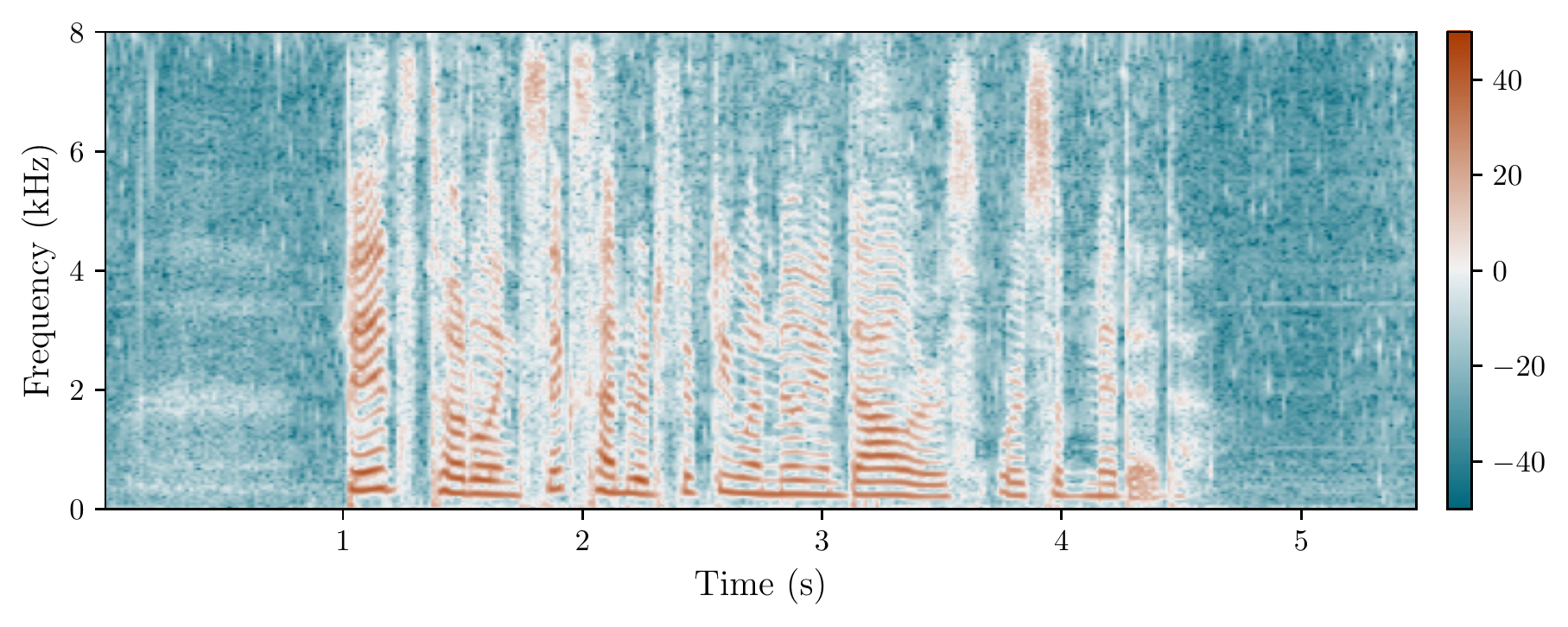}
  \caption{Adversarial Example against \kaldi}
  \label{fig:spectogram-static-ae}
  \end{subfigure}
  \begin{subfigure}{0.49\textwidth}
  \centering
  \includegraphics[trim=0 5 0 0, clip, width=\textwidth]{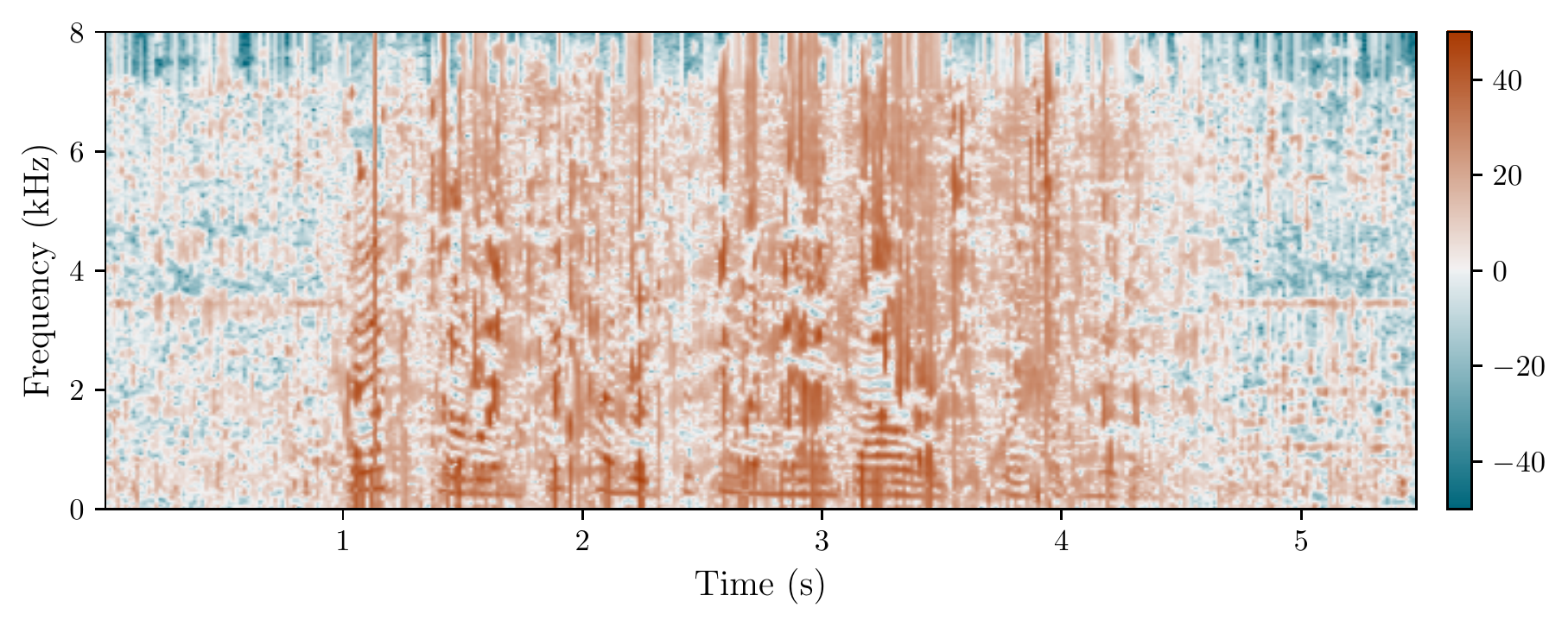}
  \caption{Adversarial Example against \tool ($\margin = 12$)}
  \label{fig:spectogram-adaptive-ae}
  \end{subfigure}
  \hfill
  \begin{subfigure}{0.49\textwidth}
  \centering
  \includegraphics[trim=0 5 0 0, clip, width=\textwidth]{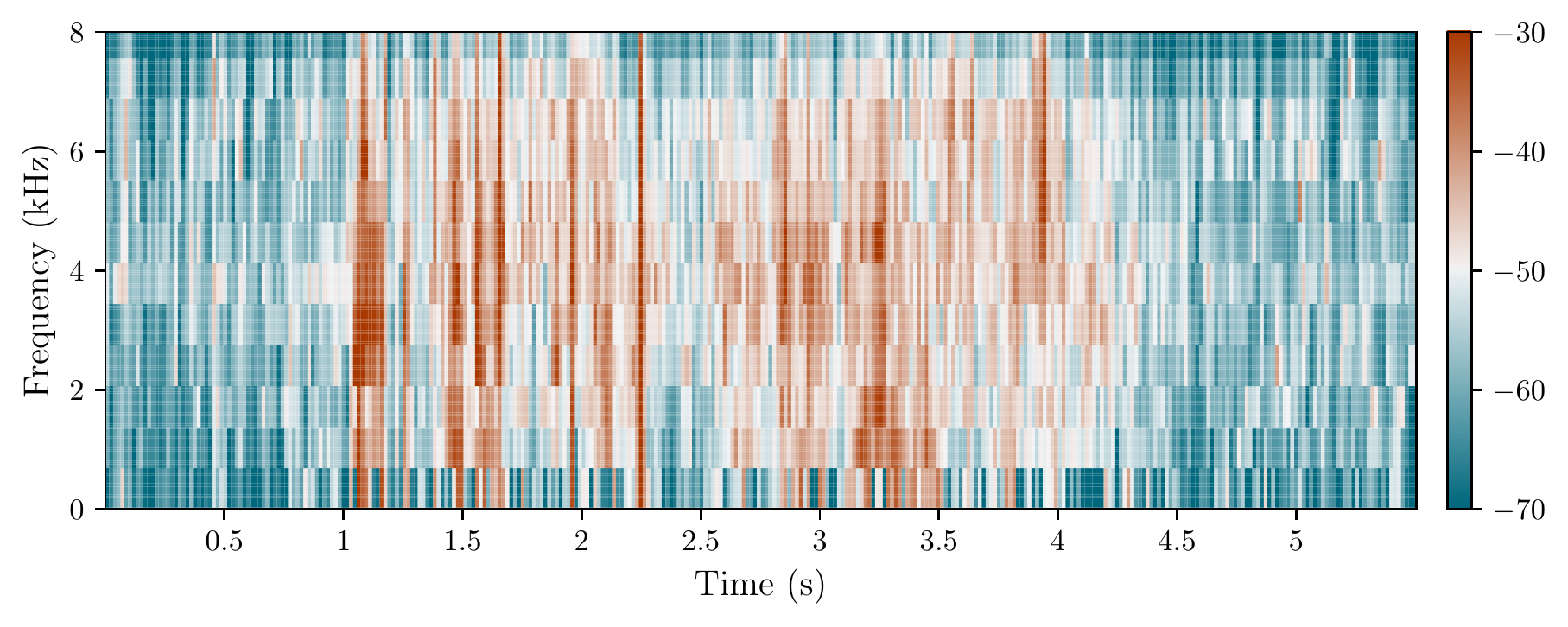}
  \caption{Hearing Thresholds}
  \label{fig:thresholds}
  \end{subfigure}
  \caption{\textbf{Spectrograms of adversarial examples.} Figure~\ref{fig:spectogram-reference} shows the unmodified signal, Figure~\ref{fig:spectogram-static-ae} depicts the baseline with an adversarial example computed against \kaldi with psychoacoustic hiding, Figure~\ref{fig:spectogram-adaptive-ae} an adversarial example computed with the adaptive attack against \tool, and Figure~\ref{fig:thresholds} shows the computed hearing thresholds for the adversarial example.}
  \label{fig:spectogram}
  \end{figure*}

%% file: includes/table-audio-content.tex
\begin{table}[t]
\centering
\footnotesize
\caption{\textbf{Number of successful Adversarial Examples (AEs) and mean Segmental Signal-to-Noise (SNRseg) ratio for non-speech audio content.} For each AE, we selected the least noisiest example, from running the attack with learning rates ($\{ 0.05, 0.1, 0.5,1.\}$). For the SNRseg we only consider successful AEs and report the difference to the baseline (\kaldi). We highlight the highest loss in bold.
}
\vspace{-0.25em}
\resizebox{1.0\columnwidth}{!}{
\begin{tabular}{@{}lrrrrrrr@{}}
\toprule
& \multicolumn{3}{c}{Birds} & & \multicolumn{3}{c}{Music}  \\ \cmidrule{2-4} \cmidrule{6-8}
\rule{0pt}{2ex}
& \multicolumn{1}{c}{\scriptsize{AEs}} & \multicolumn{1}{c}{\scriptsize{SNRseg (dB)}} & \multicolumn{1}{c}{Loss} & \quad& \multicolumn{1}{c}{\scriptsize{AEs}}  &  \multicolumn{1}{c}{\scriptsize{SNRseg (dB)}} & \multicolumn{1}{c}{Loss} \\
\midrule
\kaldi  \\
\rule{0pt}{2ex}
\; w/o hiding        & 50/50     & ~11.83 &                  &&  45/50 & ~23.26            & \\
\rule{0pt}{3ex}
\; w/\phantom{o} hiding         & ~5/50     & ~17.76 & (~+5.93)           && ~3/50  &  ~28.06 & (~+4.80)   \\
\rule{0pt}{3ex}
\tool \\
\rule{0pt}{2ex}
\; $\Phi =$ \,\phantom{1}0        & 50/50     & ~~9.58  &(~-2.25)           && 50/50    &~26.35 & (~+3.09)  \\
\rule{0pt}{3ex}
\; $\Phi =$ \,\phantom{1}6        & 31/50     & ~-2.15  &(-13.98)           && 45/50    &~16.03 & (~-7.23)  \\
\rule{0pt}{3ex}
\; $\Phi =$ \,12                  & ~5/50     & -12.25  &(\textbf{-24.08})  && ~3/50  & ~~1.94  &(\textbf{-21.32})  \\
\bottomrule
\end{tabular}}
\label{table:audio-content}
\end{table}

%% file: includes/figure-phone-rate.tex
\begin{figure}[t]
    \centering
  	\includegraphics[trim=10 10 10 10, clip, width=\columnwidth]{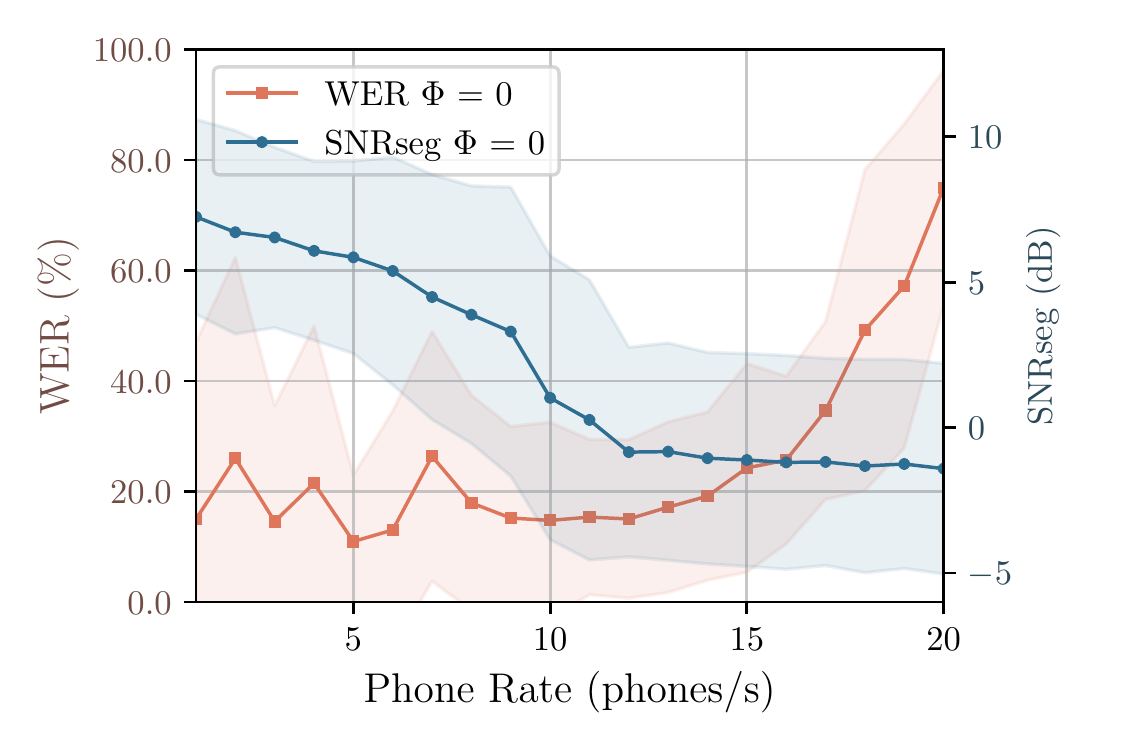}
    \caption{\textbf{Word Error Rate (WER) and Segmental Signal-to-Noise (SNRseg) ratio for different phone rates.} We report the mean and std. deviation for adversarial examples computed for targets with varying length.}
    \label{fig:phone-rates}
    \vspace{-1em}
\end{figure}

%% file: includes/table-bandpass.tex
\begin{table}[t]
\centering
\footnotesize
\caption{\textbf{Attack  for  different  cut-off  frequencies  of  the band-pass filter.} We report the number of successful adversarial examples (AEs) and the mean Segmental Signal-to-Noise (SNRseg) ratio. For the SNRseg we only consider successful AEs.}
\vspace{-0.5em}
\resizebox{1.0\columnwidth}{!}{
\begin{tabular}{@{}lrrr|rrr@{}}
\toprule
\multirow{2}{*}{Band-pass}
            & \phantom{1}300Hz- & \phantom{1}300Hz- & \phantom{1}300Hz- & \phantom{1}500Hz- & \phantom{1}500Hz- & \phantom{1}500Hz- \\
            & 7000Hz\phantom{-} & 5000Hz\phantom{-} & 3000Hz\phantom{-} & 7000Hz\phantom{-} & 5000Hz\phantom{-} & 3000Hz\phantom{-} \\
\midrule
AEs         & 18/20\phantom{\,\%} & 18/20\phantom{\,\%} & 11/20\phantom{\,\%} & 20/20\phantom{\,\%} & 17/20\phantom{\,\%} & 12/20\phantom{\,\%} \\
\rule{0pt}{2ex}
SNRseg      & 7.82\phantom{\,\%} & 7.55\phantom{\,\%} & 7.27\phantom{\,\%}  & 8.45\phantom{\,\%}  &  7.90\phantom{\,\%}  & 7.39\phantom{\,\%}   \\
\rule{0pt}{2ex}
WER      & 5.90\,\% & 5.94\,\% & 6.40\,\%  & 6.50\,\%  &  6.33\,\%  & 7.09\,\%   \\
\bottomrule
\end{tabular}}
\label{table:bandpass}
\vspace{-1em}
\end{table}

%% file: sections/05_user_study.tex
\subsection{Listening Tests}
\label{sec:user-study}
\input{includes/table-user-study}
Our goal is to make an adversarial attack noticeable by forcing modification to an audio signal into perceptible ranges.
We have used the \ac{SNRseg} as a proxy of the perceived audio quality of generated adversarial examples. However, this value can only give a rough approximation, and we are in general more interested in the judgment of human listeners. 
Specifically, we are interested to quantify if and to what extent malicious perturbations are audible to human listeners.

Therefore, we have conducted a \ac{MUSHRA} test~\cite{schinkel-13-inexperienced}, a commonly used test to assess the quality of audio stimuli.
This test allows us to get a ranking of the perceived quality of adversarial examples in comparison to an unmodified reference signal. Based on this measure, we can derive whether a participant 1) could detect any difference between an adversarial example and a clean signal (i.e., whether perturbations are audible) and, 2) obtain a subjective estimate on the amount of perceived perturbations (i.e., poorly rated samples are perceived more noisy).

\paragraph{Study Design}
\label{sec:study-design}

In a \ac{MUSHRA} test, the participants are presented with a set of differently processed audio files, the \emph{audio stimuli}.
They are asked to rate the quality of these stimuli on a scale from 0 (bad) to 100 (excellent). 
To judge whether the participants are able to distinguish between different audio conditions, a \ac{MUSHRA} test includes two additional stimuli: (i) an unaltered version of the original signal (the so-called \emph{reference}) and (ii) a worst-case version of the signal, which is created by artificial degrading the original stimulus (the so-called \emph{anchor}).
In an ideal setting, the reference should be rated best, the anchor worst.

We want to rank the perceived quality of adversarial examples computed against \tool and \kaldi.
For \tool, we select three different versions: each model uses a $200-7000\,Hz$ band-pass filter, and we vary the degree of the psychoacoustic filtering ($\margin \in \{0, 6, 12\}$).
For \kaldi, we calculate adversarial examples against the unaltered system with psychoacoustic hiding enabled (\cf~Section \ref{sec:examples}) to compare against state-of-the-art adversarial examples.

As the reference, we use the original utterance, on which the adversarial examples are based.
To be a valid comparison, we require the anchor to sound \emph{similar}, yet noisier than the adversarial examples.
Otherwise, it could be trivially identified and would not serve as a valid comparison.

Thus, we construct the anchor as follows:
For a given set, we scale and sum the noise of each of the three adversarial examples and add this sum to the original stimulus, such that 1) each noise signal contributes the same amount of energy and 2) the SNRseg of the anchor is at least 6dB lower than the SNRseg of any of the adversarial examples in the set. 

We have prepared a \ac{MUSHRA} test with six test sets based on three different audio types: two speech sample sets, two music sample sets, and two sample sets with bird sounds. 

These sets were selected among the sets of successful adversarial examples against all four models. For each set, we picked the samples whose adversarial examples produced the highest SNR (\ie the ''cleanest``) for the strongest version of \tool ($\margin = 12$).
The target text remained the same for all adversarial examples, and in all cases, the attacks were successful within $2000$ iterations.

\paragraph{Results}
To test our assumptions in the field, we have conducted a large-scale experimental study.
The G*Power 3 analysis~\cite{faul-09-statistical} identified that a sample size of 324 was needed to detect a high effect size of $\eta^2 = .50$ with sufficient power $(1-\beta >.80)$ for the main effect of univariate analyses of variance (UNIANOVA) among six experimental conditions and a significance level of $\alpha = .05$. 

We used Amazon MTurk to recruit 355 participants ($\mu_{\text{age}} = 41.61$ years, $\sigma_{\text{age}} = 10.96$; 56.60\% female). Participants were only allowed to use a computer and no mobile device. However, they were free to use headphones or speakers as long as they indicated what type of listening device was used. To filter individuals who did not meet the technical requirements needed, or did not understand or follow the instructions, we used a control question to exclude all participants who failed to distinguish the anchor from the reference correctly.

In the main part of the experiment, participants were presented with six different audio sets (2 of each: speech/bird/music), each of which contained six audio stimuli varying in sound quality. After listening to each sound, they were asked to rank the individual stimulus by its perceived sound quality. After completing of the tasks, participants answered demographic questions, were debriefed (MTurk default), and compensated with 3.00 USD. The participant required on average approximately 20 minutes to finish the~test.

In a first step, we first use an UNIANOVA to examine whether there is a significant difference between the six audio stimuli and the perceived sound quality. Our analysis reveals a significant main effect of the audio stimulus on the perceived sound quality,  $\text{\textit{F}}(5, 12780) = 8335.610$, $\text{\textit{p}} < .001$, $\eta^2 = .765$. With an alpha level of $> 1\%$ for our p-value and an effect size of $\eta^2 >.5$, our result shows a high experimental significance~\cite{richardson-11-eta}. Thus, we can conclude that \tool indeed forces adversarial perturbations into the perceptible acoustic range of human listeners.

To examine whether the effect remains stable across different audio samples and listening devices, we further conducted multiple regression analyses.
We entered the audio stimuli as our main predictors (first step) and the type of device (second step) as covariates for each analysis. Our results remain stable across all audio types. The highest predictive power was found in the \textit{speech} sets, where 82.1\% of the variance is explained by our regression model, followed by \textit{music} (76.1\%) and \textit{bird} sets (69.2\%) (see Table \ref{table:user-study} for details).
Moreover, we found a small yet significant positive coefficient for the type of device used across all audio types. This finding suggests that headphone users generally indicate higher quality rankings, potentially due to better sound perceptions. The results with listening device \emph{speaker} are presented in Figure \ref{fig:mushra}. Importantly, all results remain stable across the control variables of age, gender, and first language.

In conclusion, the results strongly support our hypothesis that \tool forces the attacker into the audible range, making the attack clearly noticeable for human listeners.

\input{includes/figure-user-study}

%% file: includes/table-user-study.tex
\begin{table}[t]
\centering
\footnotesize
\astretch{1.05}
\caption{\textbf{Regression results for perceived sound quality predicted by different audio stimuli.} The dependent variable is the quality score assigned to each audio stimulus. We trained three different models, one for each data set (speech/music/bird). Each model consists of two steps, with the first step entering the audio stimulus as a predictor and the second step entering type of device as a covariate. All models include the control variables gender, age, and language. All regressions use ordinary least squares. Cluster adjusted standard errors are indicated in parentheses. The $\text{R}^2$ values indicate the percentage of the variance of the perceived sound quality explained by the respective audio stimuli. }
\resizebox{1.0\columnwidth}{!}{
\begin{tabular}{@{}lcccccccc@{}}
\toprule
            & \multicolumn{2}{c}{Speech} &
                                    & \multicolumn{2}{c}{Music}  &
                                                            & \multicolumn{2}{c}{Bird} \\
            \cmidrule{2-3} \cmidrule{5-6} \cmidrule{8-9}
            & \scriptsize{Step 1}  & \scriptsize{Step 2} & 
                                    & \scriptsize{Step 1} & \scriptsize{Step 2} &
                                                            & \scriptsize{Step 1} & \scriptsize{Step 2} \\
\midrule
Audio      & -.905**  & -.905**    && -.871**      & -.871**  && -.830**  & -.830**      \\
 stimulus           & (.131)    & (.131)    && (.166)        & (.166)    && (.171)    & (.171)        \\
\rule{0pt}{3ex}
Device      &           & .030**   &&               & .008     &&           & .045**         \\
            &           & (.473)    &&               & (.597)   &&           & (.615)      \\
\rule{0pt}{3ex}
Controls    & \multicolumn{2}{c}{Included} && \multicolumn{2}{c}{Included} && \multicolumn{2}{c}{Included}   \\
Obs.        & \multicolumn{2}{c}{4259}
                                    && \multicolumn{2}{c}{4259}   && \multicolumn{2}{c}{4259}   \\
\rule{0pt}{3ex}
$\text{R}^2$   & .820      & .821      &&   .760    & .761      &&   .690    & .692          \\
\bottomrule
\end{tabular}}
\begin{flushleft}
\vspace{-0.5em}
\footnotesize{\textit{P}-value < 0.05 = *, \textit{P}-value < 0.01 = **}\\
\end{flushleft}
\label{table:user-study}
\vspace{-1em}
\end{table}

%% file: includes/figure-user-study.tex
\begin{figure}[t!]
  \centering
  \begin{subfigure}{1\columnwidth}
  \centering
  \includegraphics[trim=0 12 0 0, clip, width=\columnwidth]{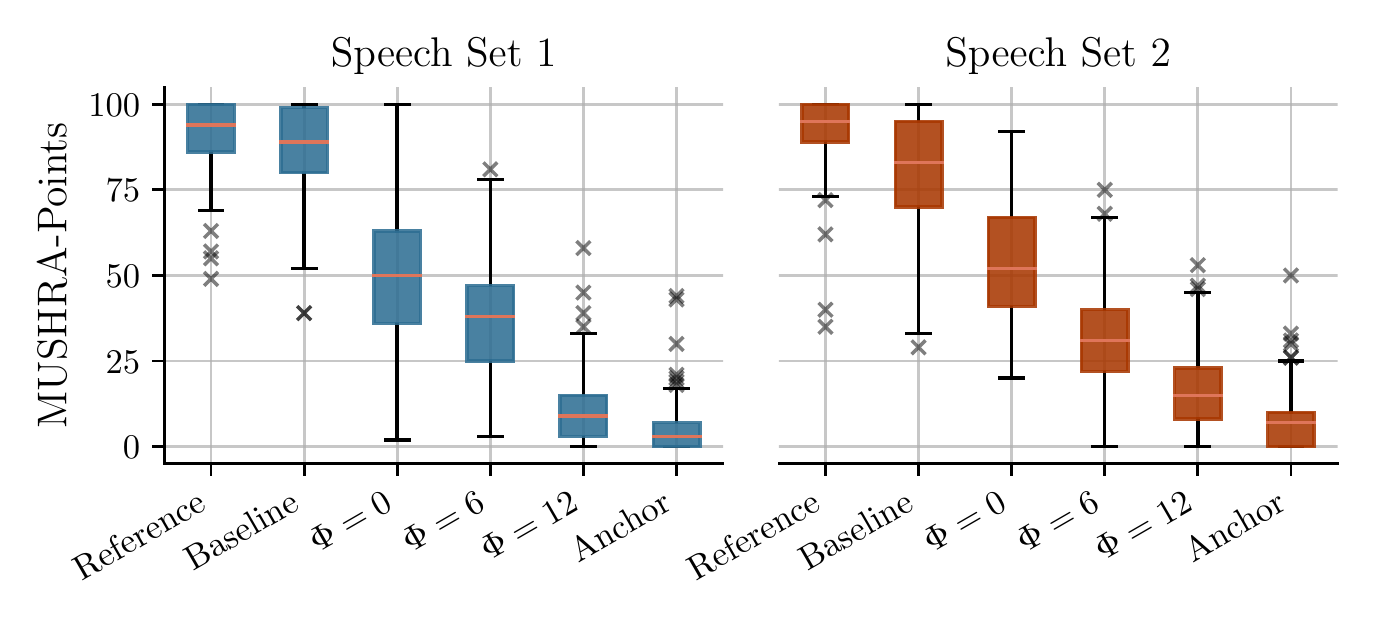}
  \caption{Speech}
  \label{fig:speech_box}
  \end{subfigure}
  \begin{subfigure}{\linewidth}
  \centering
  \includegraphics[trim=0 12 0 0, clip, width=\columnwidth]{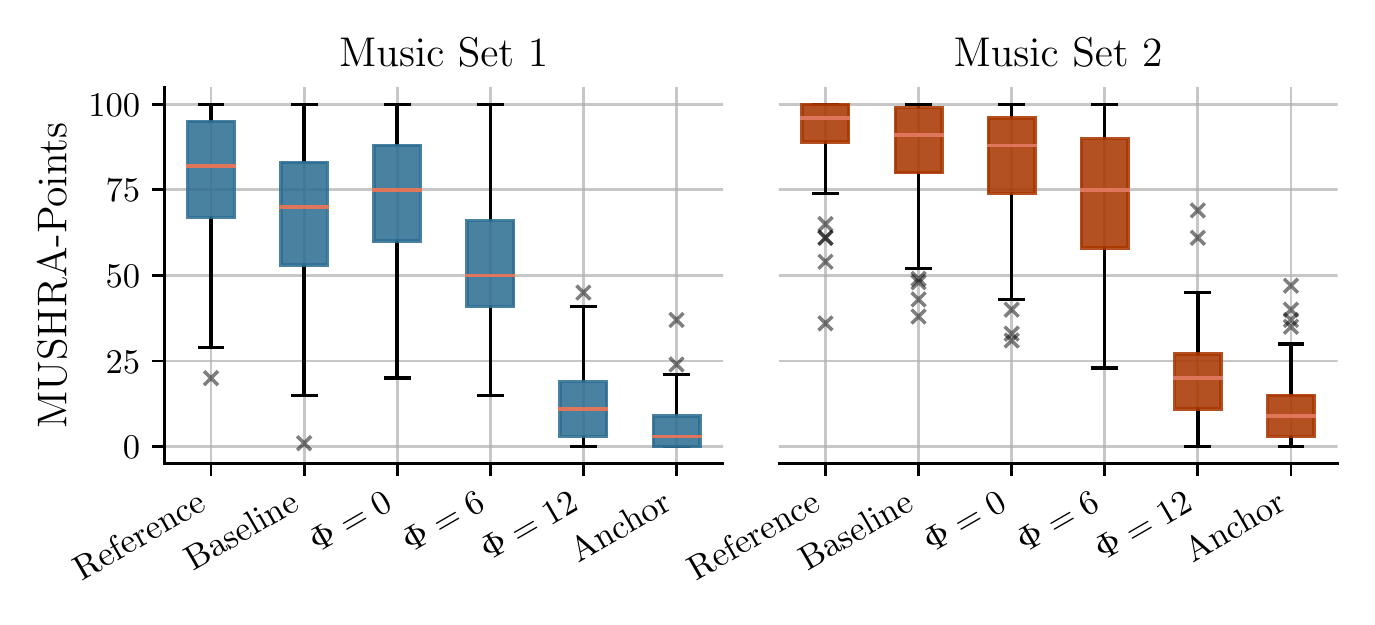}
  \caption{Music}
  \label{fig:music_box}
  \end{subfigure}
  \begin{subfigure}{\linewidth}
  \centering
  \includegraphics[trim=0 12 0 0, clip, width=\columnwidth]{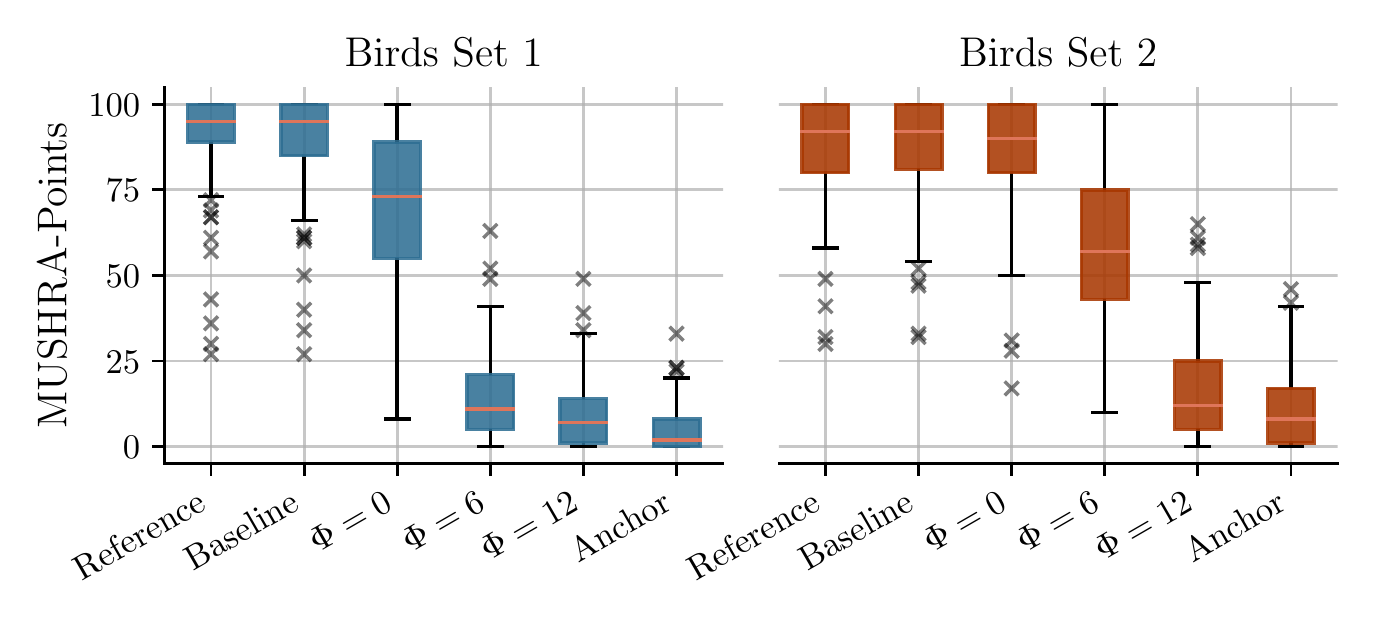}
  \caption{Birds}
  \label{fig:birds_box}
  \end{subfigure}
  \caption{\textbf{Ratings of participants with listening device \emph{speaker}.} In the user study, we tested six audio samples, divided into two samples each of spoken content, music and bird twittering.
  }
  \label{fig:mushra}
  \vspace{-1em}
\end{figure}

%% file: sections/06_related.tex
\section{Related Work}
In this section, we summarize research related to our work, surveying recent attacks and countermeasures.

\vspace{-0.5em}
\paragraph{Audio Adversarial Examples}
Carlini and Wagner~\cite{carlini-18-audio} introduced targeted audio adversarial examples for \ac{ASR} systems. For the attack, they assume a white-box attacker and use an optimization-based method to construct general adversarial examples for arbitrary target phrases against the \ac{ASR} system \textsc{DeepSpeech} \cite{hannun-14-deepspeech}.

Similarly, Sch\" onherr~\etal{}~\cite{schoenherr-19-psychoacoustics} and Yuan~\etal{}~\cite{yuan-18-commandersong} have proposed an attack against the \kaldi \cite{povey-11-kaldi} toolkit. Both assume a white-box attacker and also use optimization-based methods to find adversarial examples. Furthermore, the attack from Sch\"{o}nherr~\etal{}~\cite{schoenherr-19-psychoacoustics} can optionally compute adversarial examples that are especially unobtrusive for human listeners. 

Alzantot~\etal~\cite{alzantot-18-did} proposed a black-box attack, which does not require knowledge about the model. For this, the authors have used a genetic algorithm to create their adversarial examples for a keyword spotting system.
Khare~\etal{}~\cite{khare-19-blackbox} proposed a black-box attack based on evolutionary optimization, and also Taori~\etal{}~\cite{taori-19-targeted} presented a similar approach in their paper.

Recently, Chen~\etal~\cite{chen-20-metamorph} and Sch\"onherr~\etal~\cite{schoenherr-20-imperio} published works where they can calculate over-the-air attacks, where adversarial examples are optimized such that these remain viable if played via a loudspeaker by considering room characteristics. 

Aghakhani~\etal~\cite{aghakhani-20-venomave} presented another line of attack, namely a poisoning attack against ASR systems. In contrast to adversarial examples, these are attacks against the training set of a machine learning system, with the target to manipulate the training data s.t a model that is trained with the poisoned data set misclassifies specific inputs. 

Abdullah~\etal~\cite{abdullah-20-sok} provides a detailed overview of existing attacks in their systemization of knowledge on attacks against speech systems.

\paragraph{Countermeasures}
There is a long line of research about countermeasures against adversarial examples in general and especially in the image domain (e.\,g., \cite{metzen-17-detecting, feinman-17-detecting, carlini-17-adversarial}), but most of the proposed defenses were shown to be broken once an attacker is aware of the employed mechanism. In fact, due to the difficulty to create robust adversarial example defenses, Carlini~\etal{} proposed guidelines for the evaluation of adversarial robustness. They list all important properties of a successful countermeasure against adversarial examples~\cite{carlini-19-evaluating}.
Compared to the image domain, defenses against audio adversarial examples remained relatively unnoticed so far. For the audio domain, only a few works have investigated possible countermeasures. Moreover, these tend to focus on specific attacks and not adaptive attackers.

Ma~\etal{}~\cite{ma-19-audio-visual} describe how the correlation of audio and video streams can be used to detect adversarial examples for an audiovisual speech recognition task.
However, all of these simple approaches---while reasonable in principle---are specifically trained for a defined set of attacks, and hence an attacker can easily leverage that knowledge as demonstrated repeatedly in the image domain~\cite{carlini-17-adversarial}.

Zeng~\etal{}~\cite{zeng-18-multiversion} proposed an approach inspired by multiversion programming. Therefore, the authors combine the output of multiple \ac{ASR} systems and calculate a \emph{similarity score} between the transcriptions. If these differ too much, the input is assumed to be an adversarial example. The security of this approach relies on the property that current audio adversarial examples do not transfer between systems --- an assumption that has been already shown to be wrong in the image domain~\cite{papernot-16-transferability}.

Yang~\etal{}~\cite{yang-18-characterizing}, also utilize specific properties of the audio domain and uses the temporal dependency of the input signal. For this, they compare the transcription of the whole utterance with a segment-wise transcription of the utterance. In the case of a benign example, both transcriptions should be the same, which will not be the case for an adversarial example. This proved effective against static attacks, and the authors also construct and discussed various adaptive attacks but these were later shown to be insufficient \cite{tramer-20-adaptiveattacks}.

Besides approaches that aim to harden models against adversarial examples, there is a line of research that focuses on detecting adversarial examples: Liu and Ditzler~\cite{liu-20-detecting} utilizing quantization error of the activations of the neural network, which appear to be different for adversarial and benign audio examples. 
Däubener~\etal~\cite{daubener-20-detecting} trained neural networks capable of uncertainty quantification to train a classifier on different uncertainty measures to detect adversarial examples as outliers. Even if they trained their classifier on benign examples only, it will most likely not work for any kind of attack, especially those aware of the detection mechanism.

In contrast, our approach does not rely on detection by augmenting the entire system to become more resilient against adversarial examples. The basic principle of this has been discussed as a defense mechanism in the image domain with JPEG compression~\cite{dziugaite-16-jpg, das-19-shield} as well as in the audio domain by Carlini and Wagner~\cite{carlini-18-audio}, Rajaratnam~\etal{}~\cite{rajaratnam-18-isolated}, Andronic~\etal~\cite{andronic-20-mp3}, and Olivier~\etal\cite{olivier-21-highfrequency}. 
These approaches, however, were only used as a pre-processing step to remove semantically irrelevant parts from the input and thereby destroy adversarial perturbations added by (static) attackers. In contrast, we aim to train an ASR system that uses the same information set as the human auditory systems. Consequently, adversarial examples computed against this system are also restricted to this set, and an attack cannot be hidden in inaudible ranges. Similar to the referenced approaches, we rely on psychoacoustics and baseband filtering. However, we do not solely employ this as a pre-processing step but train a new system with our augmentation data (i.e., removing imperceptible information from the training set).
This allows us to not simply destroy adversarial perturbations but rather confine the available attack surface.

%% file: sections/07_discussion.tex
\section{Discussion}
We have shown how we can augment an \ac{ASR} system by utilizing psychoacoustics in conjunction with a band-pass filter to effectively remove semantically irrelevant information from audio signals.
This allows us to train a hardened system that is more aligned with human perception.

\paragraph{Model Hardening}
Our results from Section~\ref{sec:benign-performance} suggest that the hardened models primarily utilize information available within audible ranges.
Specifically, we observe that models trained on the unmodified data set appear to use \emph{any} available signals and utilize information \emph{both} from audible and non-audible ranges.
This is reflected in the accuracy drop when presented with psychoacoustically filtered input (where only audible ranges are available). 
In contrast, the augmented model performs comparably well on both types of input.
Hence, the model focuses on the perceivable audible ranges and \emph{ignores} the rest.

\paragraph{Robustness of the System}
We demonstrated how we can create a more realistic attacker, which actively factors in the augmentations during the calculation of adversarial examples. In this case, however, the attack is forced into the audible range.
This makes the attack significant more perceptible --- resulting in an average \ac{SNRseg} drop of up to 24.33\,dB  for speech samples.
These results also transfer to other types of audio content (i.e., music and birds tweeting) and are further confirmed by the listening test conducted in Section~\ref{sec:user-study}. 
In summary, the results of these experiments show that an attack is clearly perceivable. Further, we find that the adversarial examples, calculated with the adaptive attack, are easily distinguishable from benign audio files by humans.

\paragraph{Implementation Choices}
In general, our augmentations can be implemented in the form of low-cost pre-processing steps with no noteworthy performance overhead. Only the model needs to be retrained from scratch. However, the cost of this could---in theory---be partially alleviated by transfer learning.
We leave this question as an interesting direction for future research.

\paragraph{Robustness-Performance Tradeoff}
The results of the adaptive attack (\cf Table~\ref{table:adaptive-attack}) show that a larger margin $\margin$ leads to stronger robustness. Specifically, for $\margin$ = 14, the attacker was unable to find \emph{any} successful adversarial example in our experiments. However, this incurs an expected robustness-performance trade-off as previous research indicates that adversarial robustness is generally correlated with a loss in accuracy~\cite{tsipras-19-robustness}.

In the case of our strong white-box attacker, we recommend a margin $\margin \ge 12$, which result in a degraded system performance by at least 1.82 percentage points in terms of the benign WER. In this case, though, we already granted the attacker many concessions: full access to the model with all parameters, ideal playback (i.e., adversarial examples are fed directly into the recognizer and are not played over-the-air), and an easy target. We chose to study our attacker in this setting as this poses the strongest class of attacks and allows us to gain meaningful insights.

In contrast to white-box attacks, black-box attack don't have direct access to the gradient and for example rely on surrogate models \cite{chen-20-devil} or generative algorithms \cite{du-20-sirenattack} to construct adversarial examples.
Therefore, adversarial examples from these attacks are typically more conspicuous and can even introduce semantic changes such that humans can perceive the hidden transcription if they are made aware of it \cite{chen-20-devil}.
Considering our augmentations, we expect that current black-box attacks are able to construct valid adversarial examples against \tool.
However, we expect these to be significantly more noisy (in comparison to the adaptive attacker) as \tool forces modifications to the signal into audible ranges regardless of the underlying attack strategy.
Especially in a realistic over-the-air setting, we suspect much higher distortions since the attacker is much more constrained.
In such a setting, a smaller $\margin$ might also already suffice.
We leave this as an interesting research direction for future work.

\paragraph{Improvement of the Attack}
The adaptive attack presented in Section \ref{sec:adaptive_attack} can successfully compute adversarial examples, except for very aggressive filtering.
While Figure~\ref{fig:attack-itr} clearly shows that the attack has converged, we were still unable to find working adversarial examples.
However, other target/input utterance combinations may still exist, for which the attack works and novel attack strategies should be studied.

\paragraph{Forcing Semantics into Adversarial Examples}
We have shown how we can force adversarial audio attacks into the audible range.
This makes them clearly perceivable.
Ultimately, the goal is to push adversarial examples towards the perceptual boundary between original and adversarial message.
Intuitively, adversarial examples should require such extensive modification that a \emph{human listener} will perceive the target transcription, \ie that the adversarial perturbation carries \emph{semantic} meaning. 
We view our work as a first successful step into that direction and leave the exploration of this strategy as an interesting question for future work.

%% file: sections/08_conclusion.tex
\section{Conclusion}

In this work, we proposed a broadly applicable design principle for \ac{ASR} systems that enables them to resemble the human auditory system more closely.
To demonstrate the principle, we implemented a prototype of our approach in a tool called \tool. More specifically, we augment \kaldi using psychoacoustic filtering in conjunction with a band-pass filter.
In several experiments, we demonstrate that our method renders our system more robust against adversarial examples, while retaining a high accuracy on benign audio input.

We have argued that an attacker can find adversarial examples for any kind of countermeasure, particularly if we assume the attack to have full white-box access to the system. 
Specifically, we have calculated adversarial examples for \tool via an adaptive attack, which leverages the full knowledge of the proposed countermeasures.
Although this attack is successful in computing adversarial examples, we show that the attack becomes much less effective. More importantly, we find that adversarial examples are of poor quality, as demonstrated by the \ac{SNRseg} and our listening test. 

In summary, we have taken the first steps towards bridging the gap between human expectations and the reality of \ac{ASR}~systems---hence taming adversarial attacks to a certain extent by robbing them of their stealth abilities.

%% file: sections/09_appendix.tex
\section{Targets}
\label{app:targets}
\begin{table}[ht!]
\vspace{-0.5em}
\caption{\textbf{Target utterances for the experiments with the adaptive attacker.} For the experiments we select 50 utterances as target with an approximate length of 5s from the WSJ speech corpus test set \texttt{eval92}.}
\centering
\resizebox{\columnwidth}{!}{
\begin{tabular}{cc|cc|cc} 
\toprule
Utterance & Length & Utterance & Length & Utterance & Length  \\
\midrule
440c0407 & 5.47s & 440c040i & 5.07s & 440c040j & 4.08s \\
441c0409 & 4.91s & 441c040c & 5.57s & 441c040l & 5.26s \\
441c040m & 5.50s & 441c040s & 4.58s & 441c040y & 4.08s \\
442c0402 & 5.14s & 442c040c & 5.69s & 442c040d & 4.80s \\
442c040h & 4.63s & 442c040k & 5.37s & 442c040w & 5.21s \\
443c0402 & 5.05s & 443c040b & 4.69s & 443c040c & 4.73s \\
443c040d & 5.54s & 443c040j & 4.61s & 443c040l & 4.23s \\
443c040p & 4.10s & 443c040v & 4.82s & 443c0417 & 4.55s \\
444c0407 & 4.76s & 444c0409 & 5.20s & 444c040i & 4.98s \\
444c040n & 5.18s & 444c040u & 4.37s & 444c040w & 5.52s \\
444c040z & 4.29s & 444c0410 & 4.16s & 445c0409 & 5.43s \\
445c040j & 4.99s & 445c040l & 5.64s & 445c040w & 4.92s \\
445c040x & 4.68s & 445c0411 & 4.34s & 446c0402 & 5.59s \\
446c040b & 4.10s & 446c040d & 5.18s & 446c040e & 4.94s \\ 
446c040f & 4.66s & 446c040o & 5.33s & 446c040p & 5.04s \\
446c040s & 5.18s & 446c040v & 4.07s & 447c040g & 4.64s \\
447c040p & 5.23s & 447c040z & 4.68s &		   &	   \\
\bottomrule
\end{tabular}
}
\label{tab:targets}
\end{table}